\documentclass[aps,prb,twocolumn,superscriptaddress]{revtex4}
\usepackage{graphicx}

\begin{document}

\title{The Energetics of Li Off-Centering in K$_{1-x}$Li$_x$TaO$_3$:
First Principles Calculations}

\author{S. A. Prosandeev$^{1,2}$, E. Cockayne$^1$, and B. P.
Burton} \affiliation{Ceramics Division, Materials Science and
Engineering Laboratory, National Institute of Standards and
Technology, Gaithersburg, Maryland 20899-8520\\ $^2$Physics
Department, Rostov State University, 5 Zorge St., 344090 Rostov on
Don, Russia}

\date{\today}

\begin{abstract}

K$_{1-x}$Li$_{x}$TaO$_3$~(KLT) solid solutions exhibit a variety
of interesting physical phenomena related to large
displacements of Li-ions from ideal perovskite A-site positions.
First-principles calculations for KLT supercells were used to
investigate these phenomena.  Lattice dynamics calculations
for KLT exhibit a Li off-centering instability. The energetics of
Li-displacements for isolated Li-ions and for Li-Li pairs up to
4th neighbors were calculated.  Interactions between nearest
neighbor Li-ions, in a Li-Li pair, strongly favor ferroelectric alignment along
the pair axis. Such Li-Li pairs can be considered
``seeds" for polar nanoclusters in KLT. Electrostriction, local
oxygen relaxation, coupling to the KT soft-mode, and interactions
with neighboring Li-ions all enhance the polarization
from Li off-centering. Calculated hopping barriers for
isolated Li-ions and for nearest neighbor Li-Li pairs are in good
agreement with Arrhenius fits to experimental dielectric data.

\end{abstract}

\maketitle

\section{Introduction}

Potassium tantalate, KTaO$_3$ (KT), is a cubic perovskite with a
quantum paraelectric ground state.\cite{Sam73} In
K$_{1-x}$Li$_{x}$TaO$_3$~ (KLT) solid solutions, Li-ions
substitute for K on perovskite A-sites. Li-ions are smaller than
K, and displace from ideal, centrosymmetric, A-sites by about 1.26~\AA~ along cubic
[0,0,1]-type vectors ($[0,0,1]_{c}$).\cite{Borsa}  Displacements of Li-ions
generate strong local dipole moments [Li-dipole(s)] which couple electrostatically to the KT
polar soft-mode.  Near neighbor (nn) Li-dipoles can also interact
to form polar nanoclusters (PNC); i.e. multicell regions in which
the local polarization direction is strongly correlated. The
complex nature of ferroelectric (FE) ordering in KLT derives from
a combination of interactions between Li-dipoles, the soft-mode,
and PNC.

In the composition range $0<x<0.02$,  KLT is a quantum
paraelectric, but for $0.02<x< 0.06$, both FE and dipole-glass
characteristics are observed.\cite{Huchli,Vug90,Kleemann} In the
temperature ($T$) interval $0<T\lesssim100$~K, the dielectric
permittivity, $\varepsilon (T)$, has a peak, and second harmonic
generation exhibits an abrupt increase, with
hysteresis.\cite{Kapphan,EJPB} The existence of randomly
distributed and randomly poled PNC was established from dielectric
measurements, birefringence and
nuclear magnetic resonance (NMR).\cite{Huchli,Kleemann,Christen,Levelut}
Neighboring PNC may
touch, but PNC-coarsening is apparently prevented by random long-range
PNC-PNC interactions.\cite{EJPB} In zero applied field, a glassy
phase is observed,\cite{Kleemann} however, a sufficient applied
field will induce FE long-range order
(LRO).\cite{Huchli,Kleemann,Kapphan} For $0.06 \lesssim  x <
0.15$, KLT undergoes a FE phase transition at T $\gtrsim 100 $K,
even in the absence of an applied field.  At
$x > 0.15$, the perovskite solid solution is no longer stable.

Dielectric measurements on KLT with $0<x<0.02$~ exhibit only one
relaxational peak in $\varepsilon(T)$, while at $x \approx 0.04$,
two peaks are observed, and they are associated with two
relaxational processes (fast and slow).\cite{Dousenau,Christen,
Toulouse,Trepakov} Temperature variations of both relaxation
processes follow Arrhenius laws; $\tau \sim \tau_0 e^{-U/k T}$.
The fast process activation energy is $U \approx$ 1000 K (86 meV),
and the slow process, at $0.02<x< 0.06$, has been reported as
$2100\,$K,\cite{Hoechli} $2400\, \mathrm{K}$,\cite{Toulouse} and
$2800\,\mathrm{K}$.\cite{Trepakov}

The fast process $1000\,\mathrm{K}$~ activation energy is believed to be the
barrier for an isolated Li-ion to hop from one $[0,0,1]_{c}$~ position
to another separated by a 90$^\circ$ angle. NMR data support this
hypothesis.\cite{Borsa} The slow process with an activation energy of 2100
$\lesssim U \lesssim$ 2800 K \cite{Hoechli,Toulouse,Trepakov} is
less well understood. On the basis of acoustic measurements it was
hypothesized \cite{Dousenau,Christen} that the slow process might
be $180^\circ$ correlated Li-Li nn pair reorientations (e.g. $z-z \rightleftharpoons
\overline{z}-\overline{z}$~ relaxation for nn separated by $a_{0}
\hat{z}$~ where $a_{0}$~ is the lattice parameter of the primitive
cubic unit cell, and $\hat{z}$~ is a unit vector in the $[0,0,z]$~
direction).

The energy spectrum and kinetics of PNC reorientation is a general
problem for FE systems that exhibit both relaxor and soft-mode
behavior (cf. \cite{Gehring}). KLT is a relatively simple model
system for such phenomena because isolated Li-ions in KLT
solutions can be regarded as randomly distributed sources of
6-state Potts-like dipole fields.\cite{6wells,Vollmayr,Sakhnenko}
Previous modeling of KLT includes: Semi-empirical shell model
calculations \cite{Exner}; Intermediate neglect of the
differential overlap (INDO) calculations \cite{Eglitis, Tupitsin};
Full potential linear muffin-tin orbital (FPLMTO) calculations
\cite{Postnikov, Tupitsin} (without ionic relaxations). All these
studies yielded potential barriers for $90^\circ$ Li-hopping that
were significantly lower than those deduced from experimental
fits.

KLT solutions with $0.016 <x< 0.05$~ exhibit a large photocurrent
below $80 \mathrm{K}$.\cite{Glinchuk,Sangalli,Galinetto} No
photocurrent has been observed in pure KT, nor in solid solutions of
KT with KNbO$_3$.  Photocurrent is usually thought to be
associated with O-hole
centers near Li-occupied A-sites \cite{Glinchuk}.  Recent INDO
calculations \cite{Tupitsin} confirmed that displacements of O$^{2-}$-ions which
are nn of Li create shallow states in the forbidden gap.

First-principles (FP) computations allow determinations of the
electronic structures and energetics as functions of ionic
coordinates, and thus provide a sound basis for understanding and
quantitatively modeling KLT.  In this work, FP density functional
theory (DFT) methods were used to calculate the energetics of
Li off-centering for isolated Li-ions, and for Li-Li pairs up to
fourth neighbors.  The energetics of $90^{\circ}$~
isolated Li-ion hopping and $180^{\circ}$~ Li-Li
nn-pair hopping were calculated. Lattice dynamics studies of KLT
demonstrate correlated motions of Li and the surrounding ions
(particularly nn O$^{2-}$). Total polarization from Li off-centering, and
individual contributions from different ions were quantified.
Electronic band structures were calculated to investigate shallow
electronic states that may promote photocurrent.

\section{Methods}
All DFT calculations were done with the Vienna {\it ab initio}
simulation package (VASP) \cite{Kresse1,Kresse2}. A plane wave
basis set for electronic wavefunctions and ultrasoft
pseudopotentials were used \cite{vanderb} in the local density
approximation (LDA) for exchange and correlation energies. VASP
computes interatomic forces and total energies for crystals, and
allows either global or constrained relaxations of internal
coordinates and/or lattice parameters. Frozen-phonon methods were
used to obtain force constants for computing lattice dynamics.
Berry phase analyses, as implemented in VASP by M. Marsman, were
used to calculate dynamical charges.

Investigating the effects of Li-ions in KLT with low Li
concentration requires large supercells, {\it e.g.} (Fig. 1): (a)
KT40, a 40 atom supercell of pure KTaO$_3$; (b) KLT40, a 40 atom
supercell with basis vectors $[0,0,2]_{c}$, $[0,2,0]_{c}$ and
$[2,0,0]_{c}$ (units of $a_{0}$, $x$ = 0.125); (c) KLT80$a$~ an 80
atom supercell with basis vectors $[2,2,0]_{c}$, $[2,0,2]_{c}$,
and $[0,2,2]_{c}$ and $x$ = 0.0625; (d) KLT80$b$, same as
KLT80$a$~ but with two Li per supercell, $x$ = 0.125, separated by
$a_{0} \hat{z}$; (e, f, g) KLT80$c$, KLT80$d$, KLT80$e$, same as
KLT80b, but with two Li separated by $a_{0} (\hat{x}+\hat{y})$,
$a_{0} (\hat{x}+\hat{y}+\hat{z})$ and $2 a_{0} \hat{z}$,
respectively. For KT40 and KLT40, a 4$\times$4$\times$4
Monkhorst-Pack $k$-point grid was used and for KLT80a, etc., a
2$\times$2$\times$2 grid. Results for KLT80a are negligibly
different with a 4$\times$4$\times$4 Monkhorst-Pack grid.

This computational scheme was tested on pure KT with 5 atoms per
cell. The VASP lattice constant $a$ = 3.96 \AA~ agrees with
previous LDA computations\cite{Singh}, and is smaller than the
experimental values (between 3.987~\AA~and 3.989~\AA~at room
temperature\cite{Vou51,Sam73,Shirane,Zhurova}). The calculated
bulk modulus for a KT40 supercell with 4$\times$4$\times$4
$k$-point grid is 2.0 Mbar, which is lower than the
experimental\cite{Sam73} value 2.3 Mbar. A first principles LDA
result for a 5-ion primitive cell with an 8$\times$8$\times$8
$k$-point grid is 2.1 Mbar \cite{Singh}, but a full-potential
linear muffin-tin orbital calculation, primitive cell and
8$\times$8$\times$8 $k$-point grid yielded,\cite{Postn_PRB} 2.3
Mbar. Hence, ultrasoft pseudopentials provide reasonable accuracy
compared with other {\it ab initio} calculations.

\section{results}

\subsection{Reference Structures}\label{instability}

To study Li off-centering, one needs a reference structure in
which Li occupies an ideal centrosymmetric A-site. A 40-atom
reference structure, KLT40$_{\rm Ref}$, is obtained by placing all
ions of KLT40 on ideal perovskite positions and optimizing lattice
parameters and internal coordinates (except Li), subject to a
cubic symmetry constraint. The resulting cell has $a = 2 \times
3.956$ \AA. Crystallographic data for KLT40$_{\rm Ref}$, are
tabulated in Table~\ref{kltss.tab}. Because Li is smaller than K,
there is a volume contraction around Li such that nn Ta ions
displace by 0.010 \AA~towards Li, and nn O ions displace by 0.024
\AA~ towards Li.

\begin{table}[!htbp]
\caption{Calculated structure for
KLT40$_{Ref}$:KLT40 with Li occupying ideal A-site
positions, and the structure relaxed, subject to a cubic symmetry constraint.
Space group $Pm\overline{3}m$; $a = 2 \times 3.9559$~\AA.}
\begin{tabular}{c|c|ccc}
 ion & Wyckoff position & x & y & z   \\
\hline
 K$_1$  &  1a &  0.0000 & 0.0000 & 0.0000 \\
 K$_2$  &  3d &  0.0000 & 0.0000 & 0.5000 \\
 K$_3$  &  3c &  0.5000 & 0.5000 & 0.0000 \\
Li      &  1b &  0.5000 & 0.5000 & 0.5000 \\
Ta      &  8g &  0.2508 & 0.2508 & 0.2508 \\
 O$_1$  & 12i &  0.2509 & 0.2509 & 0.0000 \\
 O$_2$  & 12j &  0.2522 & 0.2522 & 0.5000
\label{klt40ref.tab}
 \end{tabular} \label{kltss.tab} \end{table}

A series of KLT40$_{\rm Ref}$~ frozen phonon calculations, with each
symmetry-independent ion displaced in turn, was used to compute
force constants, construct the dynamical matrix, and compute
zone-center normal mode frequencies and eigenvectors. Symmetry
analysis of KLT40$_{\rm Ref}$ zone-center phonons yields
$3 A_{1g} + 2 A_{2g} + 3 A_{2u} + 5 E_g
+ 3 E_u + 5 F_{1g} + 14 F_{1u} + 6 F_{2g} + 7 F_{2u}$. Normal mode
frequencies are listed in Table~\ref{kltphon.tbl}.  One zone-center
instability is found for KLT40$_{\rm Ref}$, with symmetry $F_{1u}$.
Thus, as expected, the high-symmetry reference structure is dynamically
unstable.  Table~\ref{kltssev.tab}, gives the $F_{1u}$~ instability
eigenvector,  which is dominated by a Li off-centering $[0,0,d]_{c}$~
displacement.

\begin{table}[!htbp]
 \caption{Computed zone-center normal mode symmetries and
frequencies ($\nu$~ in cm$^{-1}$) for KLT40$_{\rm Ref}$
(Table~\ref{klt40ref.tab}).  $F_{1u}$~ frequencies are for
transverse optical modes, except for the zero-frequency acoustic
triplet. Also listed are relative amplitudes $a_{rel}$ for each
mode in the fully relaxed tetragonal KLT40 structure.  The
relaxation is dominated by the eigenvector of the 172 $i$~
cm$^{-1}$ instability.  In addition to the ferroelectric
distortion itself which has $F_{1u}$ symmetry, there are secondary
distortions of $A_{1g}$- and $E_g$-type.}

\begin{tabular}{ccc|ccc|ccc}
symm. & $\nu$ & $a_{rel}$ & symm. & $\nu$ & $a_{rel}$ & symm. &
$\nu$ & $a_{rel}$ \\ \hline
 $A_{1g}$ & 211 & 0.0906 & $F_{1g}$ &70     & 0.0000 &
 $F_{2g}$ &  63 & 0.0000 \\ $A_{1g}$ & 418 &0.0877 & $F_{1g}$ & 205
 & 0.0000 & $F_{2g}$ & 178 & 0.0000 \\
 $A_{1g}$ & 445 & 0.0504 & $F_{1g}$ & 338     & 0.0000 & $F_{2g}$ &
262 & 0.0000 \\
         &     &        & $F_{1g}$ & 527     & 0.0000 & $F_{2g}$ & 327 & 0.0000 \\
$A_{2g}$ & 167 & 0.0000 & $F_{1g}$ & 611     & 0.0000 & $F_{2g}$ &
534 & 0.0000 \\ $A_{2g}$ & 285 & 0.0000 &          &         & &
$F_{2g}$ & 910 & 0.0000 \\
         &     &        & $F_{1u}$ & 172 $i$ & 2.9599 &          &     &        \\
$A_{2u}$ &  58 & 0.0000 & $F_{1u}$ &   0     & 0.0000 & $F_{2u}$ &
129 & 0.0000 \\ $A_{2u}$ & 507 & 0.0000 & $F_{1u}$ & 120     &
0.3047 & $F_{2u}$ & 159 & 0.0000 \\ $A_{2u}$ & 945 & 0.0000 &
$F_{1u}$ & 164     & 0.0596 & $F_{2u}$ & 178 & 0.0000 \\
         &     &        & $F_{1u}$ & 177     & 0.0080 & $F_{2u}$ & 243 & 0.0000 \\
$E_{g}$  & 125 & 0.6230 & $F_{1u}$ & 187     & 0.2049 & $F_{2u}$ &
256 & 0.0000 \\ $E_{g}$  & 213 & 0.1954 & $F_{1u}$ & 200     &
0.0082 & $F_{2u}$ & 331 & 0.0000 \\ $E_{g}$  & 279 & 0.2125 &
$F_{1u}$ & 205     & 0.2038 & $F_{2u}$ & 349 & 0.0000 \\ $E_{g}$ &
422 & 0.1413 & $F_{1u}$ & 244     & 0.0961 &          &     &
\\ $E_{g}$  & 448 & 0.1241 & $F_{1u}$ & 336     & 0.0901 &
&     &        \\
         &     &        & $F_{1u}$ & 364     & 0.0041 &          &     &        \\
$E_{u}$  &  69 & 0.0000 & $F_{1u}$ & 432     & 0.0373 & &     &
\\ $E_{u}$  & 495 & 0.0000 & $F_{1u}$ & 555     & 0.0378 &
&     &        \\ $E_{u}$  & 577 & 0.0000 & $F_{1u}$ & 862     &
0.0075 &          &     &        \\
 \end{tabular} \label{kltphon.tbl} \end{table}

\begin{table}[!htbp]
\caption{Dynamical matrix eigenvector for the $z$-polarized 172
$i$~ cm$^{-1}$~ mode.  Ionic labels correspond to those in
table~\ref{kltss.tab}.}
\begin{tabular}{c|ccc|ccc}
 ion & x & y & z & $e_x$ & $e_y$ & $e_z$ \\
\hline
 K$_1$  & 0.0000 & 0.0000 & 0.0000 &   0.0000 &   0.0000 &   0.0206  \\ 
 K$_2$  & 0.0000 & 0.0000 & 5.0000 &   0.0000 &   0.0000 & --0.0374  \\ 
 K$_2$  & 0.5000 & 0.0000 & 0.0000 &   0.0000 &   0.0000 &   0.0036  \\ 
 K$_3$  & 0.5000 & 0.5000 & 0.0000 &   0.0000 &   0.0000 &   0.0507  \\ 
 K$_3$  & 0.0000 & 0.5000 & 0.5000 &   0.0000 &   0.0000 & --0.0350  \\ 
Li      & 0.5000 & 0.5000 & 0.5000 &   0.0000 &   0.0000 &   0.9252  \\ 
Ta      & 0.2508 & 0.2508 & 0.2508 &   0.0300 &   0.0300 &   0.0252  \\ 
 O$_1$  & 0.2509 & 0.2509 & 0.0000 &   0.0000 &   0.0000 & --0.0509  \\ 
 O$_1$  & 0.0000 & 0.2509 & 0.2509 & --0.0187 &   0.0000 & --0.0301  \\ 
 O$_2$  & 0.2522 & 0.2522 & 0.5000 &   0.0000 &   0.0000 & --0.0242  \\ 
 O$_2$  & 0.5000 & 0.2522 & 0.2522 &   0.0000 & --0.0649 & --0.0875  \\ 
 \end{tabular} \label{kltssev.tab} \end{table}

For comparison, the force constants of a 40-atom cell of pure
KTaO$_3$ (KT40) were calculated at the same lattice parameter.
The TO normal mode results are listed in Table~\ref{ktphon.tbl}.
Force constants for KT40 and KLT40$_{\rm Ref}$ are nearly
identical. The only interatomic force constants that change by more
than 0.41 eV/\AA$^2$ are those involving the 12 O-ions that are
nn of Li. For example, the radial interionic force constant between
an A-ion and a nn-O changes sign from $-0.21$ eV/\AA$^2$~ in KT
to $+0.74$ eV/\AA$^2$~ in KLT40$_{\rm Ref}$ when K is replaced by Li.
The positive sign for this term in KLT40$_{\rm Ref}$ implies that Li-motion
in opposition to its nn O-ions is energetically favorable. Thus, the
single lattice instability in KLT40$_{\rm Ref}$ is dominated by
Li-motion (85\% of the dynamical matrix eigenvector) opposite its
nn O (12\% of the dynamical matrix eigenvector). Because KLT40$_{\rm Ref}$~ and
KT40 have such similar force constant matrices, their normal mode
spectra are very similar. Note that each TO mode in KT40 has a corresponding
TO mode of very similar frequency in KLT40$_{\rm Ref}$.\cite{foot1}

\begin{table}[!htbp] \caption{Computed TO normal mode frequencies
($\nu$~ in cm$^{-1}$) and dynamical matrix eigenvectors for
KTaO$_3$ at $a$ = 3.956~\AA.}
\begin{tabular}{c|cccc}
$\nu$ & $u_{K}$ & $u_{Ta}$ & $u_{O\parallel}$ & $u_{O\perp}$ \\
\hline
115 & --0.2989 &   0.5419 & --0.3937 & --0.4807 \\
205 &   0.8745 & --0.1737 & --0.2401 & --0.2715 \\
555  &  0.0012 &   0.0341 & --0.8531 &   0.3682
\end{tabular} \label{ktphon.tbl} \end{table}

\subsection{Fully Relaxed Structures}\label{relaxed}

Full ionic relaxation in KLT40 indicates that (consistent with
previous unrelaxed computations \cite{Postnikov,Tupitsin}) maximum
energy reduction occurs when Li-ions are displaced to
$[0,0,d]_{c}$~ (Figs. 2-3). The calculated equilibrium
Li-displacement\cite{foot2} is $d$=1.009~\AA~ (Table
\ref{POSCAR}), whereas fits to experimental data yield
\cite{Borsa} $d=1.26$ \AA. For comparison, the curves obtained
when all ions, except for Li, are fixed at ideal perovskite
positions, are also shown in Fig. 3. Clearly, the depth of the
well and the magnitude of Li-displacement are sensitive to
relaxations of the surrounding ions, consistent with earlier
results. \cite{Eglitis}

The largest sympathetic distortion associated with
Li-displacements involve four nn O-ions (see Table \ref{POSCAR}
and Fig. 2), which are displaced towards Li by 0.109 \AA. Similar
correlated Li- and O-displacements occur in LiNbO$_{3}$~ and
LiTaO$_{3}$~\cite{Cohen}. In each case, as Li displaces, the
oxygen atoms that Li approaches move toward the Li, and those left
behind back away, as might be expected from electrostatics. The
magnitude of Li displacement is about 0.3~\AA~in LiTaO$_{3}$~
\cite{Cohen} or 0.4~\AA~with full geometry
optimization,\cite{PostnLiTaO3} i.e., much smaller than Li in KLT.
Note, however, that LiTaO$_3$ and LiNbO$_3$ have the LiNbO$_{3}$~
structure type, a derivative of the corundum structure, not the
perovskite type, and that the displacements patterns are therefore
not directly comparable.

\begin{table}[!hbp]
\caption{Computed fully relaxed structure of KLT40. Space group
P4mm; $a = 2 \times 3.9565$~\AA; $c = 2 \times 3.9638$~\AA; $c/a
\approx 1.002$. Deviations are relative to positions in
KLT40$_{\rm Ref}$. The origin is chosen so that both structures
have the same center of mass. Wyckoff positions (pos.) are used.}

\begin{tabular}{c|c|ccc|ccc}
 ion & pos. & x & y & z & $\delta x,$~ \AA & $\delta y$, \AA
&$\delta z$, \AA
\\ \hline
    K$_1$  & 1a &  0.0000 & 0.0000 & 0.0014 &   0.000 &  0.000 &  0.011  
 \\ K$_2$  & 2c &  0.5000 & 0.0000 & 0.9996 &   0.000 &  0.000 & --0.003  
 \\ K$_3$  & 2c &  0.5000 & 0.0000 & 0.4963 &   0.000 &  0.000 & --0.029  
 \\ K$_4$  & 1a &  0.0000 & 0.0000 & 0.4952 &   0.000 &  0.000 & --0.038  
 \\ K$_5$  & 1b &  0.5000 & 0.5000 & 0.0074 &   0.000 &  0.000 &  0.058  
 \\ Li     & 1b &  0.5000 & 0.5000 & 0.6273 &   0.000 &  0.000 &  1.009
 \\ Ta$_1$ & 4d &  0.2513 & 0.2513 & 0.2526 &   0.004 &  0.004 &  0.015
 \\ Ta$_2$ & 4d &  0.2494 & 0.2494 & 0.7499 & --0.011 &--0.011 &  0.005
 \\ O$_1$ & 4d &   0.2515 & 0.2515 & 0.9935 &   0.005 &  0.005 & --0.051  
 \\ O$_2$ & 4d &   0.2474 & 0.2474 & 0.4957 & --0.038 &--0.038 & --0.034  
 \\ O$_3$ & 4e &   0.0000 & 0.2500 & 0.2436 &   0.000 &--0.007 & --0.058  
 \\ O$_4$ & 4e &   0.0000 & 0.2533 & 0.7457 &   0.000 &  0.019 & --0.027  
 \\ O$_5$ & 4f &   0.5000 & 0.2482 & 0.2442 &   0.000 &--0.031 & --0.063  
 \\ O$_6$ & 4f &   0.5000 & 0.2589 & 0.7379 &   0.000 &  0.053 & --0.079  
 \end{tabular} \label{POSCAR} \end{table}

    The results listed in Table \ref{POSCAR} indicate that Ta-ions are
generally displaced in the same direction as Li, while O-ions shift in the
opposite direction.  This is consistent with the view that Li-displacement
leads to freezing of a soft-mode polarization fluctuation that has a large
correlation radius.

   A breakdown of the displacement pattern into normal mode coordinates for
KLT40$_{\rm Ref}$~ is shown in Table~\ref{kltphon.tbl}. Anharmonic coupling
leads to some contributions from higher frequency modes, but otherwise,
the displacement pattern is dominated by the eigenvector of the structural
instability.  Vibrational frequencies for fully relaxed KLT40 were also
calculated, and symmetry analysis of the fully relaxed tetragonal structure
indicates that the phonon spectrum is
22 $A_1 + 8 A_2 + 14 B_1 + 12 B_2 + 32 E$.  Modes belonging to irreducible
representations $A_1$~ and $E$~ are infrared active (except for the
$A_1 + E$~ acoustic set). The frequencies of infrared active modes are
listed in Tables~\ref{klte.tb2}. Each column represents the (summed)
projection (in percent) of the dynamical matrix eigenvector of
relaxed KLT40 onto one (or more) eigenvectors of KLT40$_{\rm Ref}$.  If
two eigenvectors were identical, the entry would be 100.

\begin{table}[!htbp]
\caption{Computed TO normal mode frequencies ($\nu$~ in cm$^{-1}$)
for modes of symmetry $E$~ (left) and $A_1$~ (right) in fully
relaxed KLT40.  Also shown is the squared projection (in percent)
of the eigenvector of each mode onto an eigenvector, or set of
eigenvectors, of the dynamical matrix for KLT40$_{\rm Ref}$. (1)
The 172$i$~ cm$^{-1}$~ instability, (2) the 120 cm$^{-1}$~ mode
similar to the KT40 soft-mode. (3) All other polar eigenvectors
and (4) All nonpolar eigenvectors.}
\begin{tabular}{c|cccc|c|cccc}
 $\nu$ & (1) & (2) & (3) & (4)&  $\nu$ & (1) & (2) & (3) & (4) \\
 \hline
  66 &  0.3 &  0.6 &  0.1 & 99.0 & 147 &  0.1 &  0.4 & 17.2 & 82.4  \\
  79 &  0.1 &  0.1 &  0.1 & 99.7 & 163 &  0.1 & 87.9 &  8.4 &  3.6  \\
 124 &  7.2 & 89.0 &  1.2 &  2.7 & 169 &  0.8 &  3.1 & 84.1 & 12.0  \\
 147 &  0.0 &  0.9 &  0.8 & 98.2 & 176 &  9.9 &  0.1 & 85.6 &  4.3  \\
 159 &  0.0 &  0.2 & 21.7 & 78.1 & 179 &  9.4 &  0.1 & 84.2 &  6.3  \\
 160 &  0.6 &  0.6 & 77.6 & 21.3 & 199 &  4.4 &  0.1 & 94.2 &  1.2  \\
 176 &  0.1 &  0.4 & 23.2 & 76.3 & 205 &  7.0 &  3.1 & 59.9 & 30.0  \\
 178 &  0.0 &  0.1 & 80.5 & 19.4 & 210 &  2.8 &  1.5 & 14.8 & 80.8  \\
 187 &  1.3 &  0.9 & 91.4 &  6.4 & 215 &  0.2 &  1.9 & 35.5 & 62.5  \\
 199 &  0.5 &  0.6 & 91.6 &  7.3 & 235 & 22.4 &  0.2 & 60.7 & 16.6  \\
 199 &  0.5 &  0.3 & 51.7 & 47.5 & 262 & 30.8 &  1.0 & 57.5 & 10.6  \\
 206 &  3.3 &  1.1 & 46.6 & 49.1 & 290 &  5.0 &  0.2 &  2.2 & 92.7  \\
 224 &  0.2 &  0.1 & 10.2 & 89.5 & 337 &  2.2 &  0.1 & 96.7 &  1.1  \\
 239 &  7.3 &  0.6 & 70.7 & 21.4 & 362 &  0.0 &  0.0 & 96.7 &  3.3  \\
 247 &  1.6 &  0.1 &  6.2 & 92.2 & 417 &  0.0 &  0.0 & 24.4 & 75.6  \\
 252 &  6.1 &  0.3 & 20.8 & 72.8 & 428 &  2.8 &  0.0 &  0.9 & 96.3  \\
 262 &  0.0 &  0.0 &  0.7 & 99.3 & 444 &  0.4 &  0.0 & 63.9 & 35.7  \\
 307 & 55.1 &  3.4 & 17.7 & 23.9 & 449 &  1.4 &  0.0 &  0.8 & 97.7  \\
 332 &  0.8 &  0.1 &  0.8 & 98.3 & 459 &  0.1 &  0.0 & 13.3 & 86.6  \\
 335 &  0.8 &  0.0 & 14.6 & 84.6 & 565 &  0.1 &  0.2 & 99.5 &  0.2  \\
 336 &  0.2 &  0.0 & 63.6 & 36.3 & 864 &  0.0 &  0.0 & 99.5 &  0.4  \\
 342 &  0.0 &  0.0 & 25.3 & 74.7 \\
 351 &  0.3 &  0.0 &  5.6 & 94.2 \\
 378 & 13.5 &  0.6 & 77.2 &  8.7 \\
 432 &  0.0 &  0.0 & 99.7 &  0.3 \\
 528 &  0.1 &  0.0 & 15.1 & 84.8 \\
 542 &  0.0 &  0.0 &  3.3 & 96.7 \\
 561 &  0.1 &  0.0 & 81.7 & 18.3 \\
 622 &  0.0 &  0.0 &  0.7 & 99.2 \\
 857 &  0.0 &  0.0 & 99.6 &  0.4 \\
 909 &  0.1 &  0.0 &  0.3 & 99.6
 \end{tabular} \label{klte.tb2} \end{table}

The KT soft-mode frequency (at equilibrium lattice constant),
120 cm$^{-1}$, splits into two modes in fully
relaxed KLT40: $E$~ (124 cm$^{-1}$), and $A_1$~ (163 cm$^{-1}$).
Hence, the averaged frequency for these modes is increased by
Li-displacement as indicated by experimental data for hardening of
the soft-mode in KLT \cite{Vogt}. However, calculated hardening of
the soft mode is only significant for the component of the KT
soft-mode triplet that is polarized in the Li-displacement direction.

To study $x$-dependence of the soft mode splitting, lattice
dynamics were calculated for KT4O ($x = 0$), KLT80a ($x =
0.0625$), and KLT40 ($x = 0.125$), all with fixed $a = 3.9879$\AA~
(the experimental value\cite{Zhurova} for $x = 0.05$). Results for
the soft mode splitting are shown in Table~\ref{split.tab}. The
calculated splitting increases with $x$. TO splitting in the
ferroelectric phase of KLT has been observed experimentally by
Raman spectroscopy\cite{Prater81} and by neutron
scattering\cite{Klein_JPCM_96}.  The measured
splitting\cite{Klein_JPCM_96} in tetragonal KLT with $x = 0.05$~
is given in Table~\ref{split.tab}. The experimental splitting is
less than that obtained from FP results, owing to the use of
periodic boundary conditions in FP calculations, which fixes the
arrangement of the Li-sites and artificially aligns Li-dipoles
from one supercell to the next (cf. Ref. [\cite{finite_q}]).

\begin{table}[!hbp]
\caption{$F_{1u} \rightarrow E + A_1$~ splitting of the
KLT TO soft mode for various supercells at fixed
$a = 3.9879$\AA.  Experimental results are
given for $x = 0.05$.  Frequencies are in cm$^{-1}$.}

\begin{tabular}{c|c|c|c|c}
 supercell & $x$ & $\nu_{TO}$~ ($E$) & $\nu_{TO}$~ ($A_1$) & splitting \\ \hline
  KT40     & 0 & (76)  &  (76)   &  0   \\
  KLT80$a$   & 0.0625 & 79  & 139  &  60  \\
  KLT40    & 0.125  & 85  & 197  & 112  \\
  Experiment\cite{Klein_JPCM_96} & 0.05  & {\it c.} 50 & {\it c.} 80 &
    {\it c.} 30 \\
\end{tabular} \label{split.tab} \end{table}

The Li-dominated instability of KLT40$_{\rm Ref}$~
mixes with several other modes; its weighted
average frequency is about 260 cm$^{-1}$~ in the polar direction
and 330 cm$^{-1}$~ in transverse directions.

To investigate the Li-isotope effect in KLT, phonon frequencies in
Table~\ref{klte.tb2} were recalculated using $^6$Li ($m = 6.015$~
amu) instead of natural Li ($m = 6.941$~ amu). Individual mode
frequencies increase when $^6$Li is used, but the $A_1$~ soft-mode
component frequency increases by only about 0.04 cm$^{-1}$, owing
to the small participation of Li in this mode. The $E$-mode
frequency increase is an order of magnitude smaller.

Computed, averaged, infrared reflectivity spectra for poled
KLT40 and KT at the equilibrium lattice constant are compared in
Fig. 4. The two spectra are generally similar, but in KLT40,
KT-bands are split by symmetry-breaking Li-displacements.

\subsection{Isolated Li-Ions}

Li-ions can jump between nn $[0,0,d]_{c}$~ wells via
$[1,1,0]_{c}$~ saddle points. Fig. 5 shows the computed Li
potential between nn Li-wells; e.g. $[0,0,d]_{c}$~ and
$[0,d,0]_{c}$~ via a minimum energy $[0,1,1]_{c}$~ saddle. Li-ions
were moved along a straight line between the wells. The energetics
of Li motion, with all non-Li ions fixed in the positions that
they occupy when Li is in its equilibrium off-center state, are
contrasted with the energetics for Li motion in conjunction with
correlated relaxations of non-Li ions. Note that these frozen
coordinates differ from those in Fig. 2 where all non-Li ions
occupy positions that are consistent with a center of symmetry.
Clearly, ionic relaxation promotes Li-hopping by reducing
potential barriers. Fully relaxed calculations yield reasonable
agreement with results obtained from Arrhenius fits to NMR data
\cite{Borsa}. The fixed-ion potential barrier is $\sim$103 meV
(1190 K). The experimental value, 86 meV (1000 K) [\cite{Huchli}]
is lower, but the value 103 meV (1200 K) [\cite{Toulouse}] is
close. A similar potential barrier value was also obtained in the
KLT80$a$~ computation. Note that computed barrier heights will be
systematically larger than values obtained by experiment if the
latter are reduced by Li-tunneling.

\subsection{Li-Li Pairs}\label{Li-pairs}

\begin{table}[!htbp]
 \caption{Fully relaxed Li-Li pair configurations in
KLT80$b$~ (no subscript), KLT80$c$~ (subscript 2), KLT80$d$~ (3)
and KLT80$e$~ (4).  Li-displacement vectors from ideal perovskite
positions are given in~\AA.}
\begin{tabular}{c|ccc|ccc|c}
 notation& $x_1$ & $y_1$ & $z_1$ & $x_2$ & $y_2$ & $z_2$  & $E
 [eV]$
 \\ \hline
 $z-z$            & 0.00  & 0.00     &   1.43  &   0.00  & 0.00     & $ 1.19$ & 0.000 \\
 $x-z$            & 1.18  & 0.00     &   0.06  & --0.01  & 0.00     & $ 1.31$ & 0.124 \\
 $x-\overline{z}$ & 1.11  & 0.00     & --0.02  &   0.00  & 0.00     & $-1.12$ & 0.221 \\
 $y-x$            & 0.00  & 1.12     & --0.01  &   1.12  & 0.00     & $ 0.01$ & 0.227 \\
 $x-x$            & 1.07  & 0.00     &   0.00  &   1.07  & 0.00     & $ 0.00$ & 0.257 \\
 $x-\overline{x}$ & 1.10  & 0.00     & --0.01  & --1.10  & 0.00     & $ 0.01$ & 0.279 \\
 $\overline{z}-z$ & 0.00  & 0.00     & --1.09  &   0.00  & 0.00     & $ 1.09$ & 0.335 \\
 $z-\overline{z}$ & 0.00  & 0.00     &   0.76  &   0.00  & 0.00     & $-0.76$ & 0.550 \\
 $x-x_2$             &  1.14 &  0.01 &--0.01 &  1.13 &  0.01 &--0.01 & 0.141 \\
 $x-z_2$             &  1.11 &  0.00 &--0.02 &  0.03 &  0.00 &  1.13 & 0.213 \\
 $x-\overline{y}_2$  &  1.11 &--0.02 &  0.01 &  0.02 &--1.11 &  0.01 & 0.216 \\
 $z-\overline{z}_2$  &  0.00 &  0.00 &  1.14 &  0.00 &--0.00 &--1.16 & 0.235 \\
 $x-y_2$             &  1.14 &  0.03 &--0.01 &  0.04 &  1.14 &--0.02 & 0.258 \\
 $z-z_2$             &  0.00 &--0.03 &  1.09 &  0.00 &--0.00 &  1.09 & 0.275 \\
 $x-\overline{x}_2$  &  1.10 &--0.02 &--0.00 &--1.18 &--0.05 &  0.01 & 0.383 \\
 $z-z_3$             &--0.01 &  0.00 &  1.14 &--0.01 &  0.00 &  1.14 & 0.138 \\
 $z-x_3$             &  0.03 &  0.01 &  1.12 &  1.12 &--0.02 &  0.03 & 0.229 \\
 $\overline{z}-z_3$  &  0.01 &--0.01 &--1.09 &  0.00 &  0.00 &  1.09 & 0.385 \\
 $z-x_4$             &--0.02 &  0.00 &  1.14 &  1.11 &  0.00 &--0.01 & 0.213 \\
 $z-z_4$             &  0.00 &  0.00 &  1.09 &  0.00 &  0.00 &  1.08 & 0.248 \\
 $z-\overline{z}_4$  &  0.00 &  0.00 &  1.10 &  0.00 &  0.00 &--1.10 & 0.257
 \end{tabular} \label{Li2energies} \end{table}

\begin{table}[!htbp]
 \caption{Excitation energies for Li-Li nn-pairs (K).}
\begin{tabular}{c|cccccccc}
 notation& $z-z$ & $x-z$ & $x-\overline{z}$
& $y-x$ & $x-x$ & $\overline{z}-z$ & $x-\overline{x}$ &
$z-\overline{z}$
 \\ \hline
  $z- z$    &   0     &       &      &      &      &      &&
  \\
  $x- z$    &   1432  &  0    &      &      &      &      &&
  \\
  $x-\overline{z}$    &   2570  &  1138 & 0    &      &     & &      &
  \\
  $y- x$    &   2627  &  1195 & 57   & 0    &      &      &&
  \\
  $x- x$    &   2990  &  1558 & 420  & 363  & 0    &      &&
  \\
  $x-\overline{x}$    &   3170  &  1738 & 600  & 543  & 180  & 0 &   &
  \\
  $\overline{z}-z$    &   3883  &  2451 & 1313  & 1253  & 893  & 713& 0&
  \\
  $z-\overline{z} $   &   6379  &  4947 & 3809 & 3752 & 3389 &  3209  & 2496
  &0
  \\
 \end{tabular} \label{Li2excitations} \end{table}

The energetics of Li-Li nn pair displacement configurations were
studied using the KLT80$a$~ supercell (Fig. 6) for all nn-pair
configurations listed in Table \ref{Li2energies}. Total energy
calculations, with all ions relaxed and $a=4$~\AA, indicate that
the lowest energy path for converting the minimum energy $z-z$~
configuration into the $\overline{z}-\overline{z}$~ configuration
is via intermediate microstates; e.g. the energy for $z-z
\rightarrow x-y$~ is 2627 K (Table \ref{Li2excitations}) within
the experimentally reported range of values, 2100-2800 K,
\cite{Hoechli,Toulouse,Trepakov} attributed to the rearrangement
of Li-Li nn-pairs. \cite{Dousenau} Barrier height computations for
two selected low-energy paths (see Fig. 7) were performed by
converting the $z-z$~ configuration into
$\overline{z}-\overline{z}$~ via intermediate states (Fig. 6).
Note that the metastable states are separated by high barriers the
lowest of which, 356 meV (4130 K), is significantly larger than
the value 2100-2800K \cite{Hoechli,Toulouse,Trepakov} obtained by
fitting an Arrhenius expression to experimental data; this
discrepancy may be explained by Li-tunneling between metastable
states that are close in energy. Ionic potentials were computed,
at fixed lattice parameters, in a two-step sequence: 1) A Li-ion
in a Li-Li nn pair was moved along a straight line between nn
equilibrium states, while other ionic coordinates (including those
of the second Li ion) were fully relaxed. 2) The second Li-ion was
moved along a straight line between nn wells while all other ionic
coordinates (including those of the first Li ion) were fully
relaxed. Simultaneous motions of both ions in a Li-Li pair were
tried, but this path required greater energy. Hence, Li dynamics
within Li-Li pairs is Glauber type.\cite{Glauber} Significantly,
the excitation energy for $z-z \rightarrow x-x$~ Li-Li nn-pair
flipping is close to the energy for $z-z \rightarrow
x-\overline{x}$~ and $z-z \rightarrow x-y$~ flipping. These
results are inconsistent with the classic dipole-dipole
interaction expression $E \propto (\vec{\mu}_1 \cdot \vec{\mu}_2)
- 3 (\vec{\mu}_1 \cdot \hat{r}_{12}) (\vec{\mu}_2 \cdot
\hat{r}_{12})$.

If the experimental observation of two relaxation processes
\cite{Toulouse,Trepakov} corresponds to flipping of isolated
Li-ions and Li-Li nn-pairs, then significant concentrations of
both must be present. Assuming a random distribution of Li-ions in
KLT, one can calculate the densities of isolated Li-ions and Li-Li
nn pairs as functions of $x$; $x(1-x)^6$, and 3$x^2(1-x)^{10}$,
respectively (Fig. 8).

The average distance between Li-Li nn-pairs is obviously larger
than the average distance between isolated Li-ions because the
concentration of Li-Li nn-pairs ($x_{Li-Li}$) is so much lower
($x_{Li-Li} \approx 3 x^2$ for small $x$). Therefore, it is
plausible to think of randomly oriented but interacting PNC that
each contain one Li-Li nn-pair as the ``seed" plus several
isolated Li-ions. Such a system has an interesting phase
diagram.\cite{6wells} Because the energy required to reorient the
polarization axis of a Li-Li nn-pair by $90^\circ$~ is large, 2990
K, PNC polar orientations resist coarsening that involves
reorientation(s) of Li-Li nn-pair axes. This constraint on
FE-percolation \cite{EJPB} may explain the observation of a
dipole-glass like state in KLT\cite{Huchli,Kleemann} with
$0<x<0.02$.

Results for supercells with two Li-ions, separated by more than
$a_{0}$, are given in Table \ref{Li2energies}. For example,
KLT80$e$~ has three symmetry-independent configurations (Fig. 9):
(1) both Li-ions are displaced in the same direction, $z-z_4$; (2)
they are displaced in opposite directions; $z-\overline{z}_4$; (3)
they are displaced in perpendicular directions, e.g. $z-x_4$. The
$z-x_4$~ configuration has lowest energy at fixed lattice
parameter. This may be a consequence of accommodating two
off-centered Li-ions, each of which produces a strong local
tetragonal distortion, into a cell in which the global strain is
constrained to be zero.

Stachiotti {\it et al.}\cite{Stachiotti_ferroel} used a shell
model to compute Li-Li interactions in KLT out to distance 6
$a_{0}$.  Their results also indicate large strain-, and nonlinear
polarization corrections, to the dipole-dipole interaction energy.
Experimental diffuse scattering data for KLT\cite{Yong} indicate
strongly correlated planar atomic displacements rather than
chain-displacements as observed in KNbO$_3$, BaTiO$_3$, KTaO$_3$~
and SrTiO$_3$. \cite{Comes,Honjo,Shirane72,Shapiro}

\subsection{Dynamical Charges and Polarization}

\begin{table}[!htbp]
\caption{Dynamical charges $Z^*_{zz}$ in the KLT40 supercell. }
\begin{tabular}{c|c|cccccc}
 Ions && x  & y  & z & $Z^*_{zz}$ & $Z^*_{xx}$ & $Z^*_{yy}$ \\
\hline
    K$_1$  & 1a &  0.0000 & 0.0000 & 0.0014 &   1.445 &   1.142 &   1.142
 \\ K$_2$  & 2c &  0.5000 & 0.0000 & 0.9996 &   1.024 &   1.180 &   1.132  
 \\ K$_3$  & 2c &  0.5000 & 0.0000 & 0.4963 &   1.014 &   1.168 &   1.122  
 \\ K$_4$  & 1a &  0.0000 & 0.0000 & 0.4952 &   0.999 &   1.146 &   1.146  
 \\ K$_5$  & 1b &  0.5000 & 0.5000 & 0.0074 &   1.402 &   1.177 &   1.177  
 \\ Li     & 1b &  0.5000 & 0.5000 & 0.6273 &   0.574 &   1.106 &   1.106

 \\ Ta$_1$ & 4d &  0.2513 & 0.2513 & 0.2526 &   8.145 &   8.745 &   8.745
 \\ Ta$_2$ & 4d &  0.2494 & 0.2494 & 0.7499 &   8.619 &   8.479 &   8.479
 \\ O$_1$ & 4d &   0.2515 & 0.2515 & 0.9935 & --6.003 & --1.685 & --1.685  
 \\ O$_2$ & 4d &   0.2474 & 0.2474 & 0.4957 & --6.232 & --1.639 & --1.639  
 \\ O$_3$ & 4e &   0.0000 & 0.2500 & 0.2436 & --1.724 & --6.382 & --1.738  
 \\ O$_4$ & 4e &   0.0000 & 0.2533 & 0.7457 & --1.661 & --6.358 & --1.619  
 \\ O$_5$ & 4f &   0.5000 & 0.2482 & 0.2442 & --1.651 & --6.662 & --1.644  
 \\ O$_6$ & 4f &   0.5000 & 0.2589 & 0.7379 & --1.617 & --6.197 & --1.790  
 \\
\end{tabular}\label{DynCh}
\end{table}

\begin{table}[!htbp]
\caption{Dynamical charges $Z^*_{zz}$, ionic displacements (in
\AA) and total dipole moment (in e\AA) in the KLT40 supercell.
Only symmetrically distinct ions are listed: for ionic coordinates
see Table \ref{POSCAR}.}
\begin{tabular}{c|ccccc}
 Ions &  Refer. & Ideal perov. & Average & Displ. &
 Tot.~dip. \\
      & struct.&positions& dyn.~ch.& [\AA] & [e\AA]\\
\hline
 K$_1$  &  1.153 & 1.152  &  1.250  &  0.011  &  0.014 \\
 K$_2$  &  1.153 & 1.151  &  1.109  &--0.003  &--0.007 \\
 K$_3$  &  1.153 & 1.153  &  1.107  &--0.029  &--0.064 \\
 K$_4$  &  1.153 & 1.153  &  1.102  &--0.038  &--0.042 \\
 K$_5$  &  1.153 & 1.124  &  1.217  &  0.058  &  0.071 \\
 all K  &        &        &  1.143  &--0.005  &--0.029 \\
 \hline
 Li     &  1.251 & 1.251  &  1.025  &  1.009  &  1.035 \\
 \hline
 Ta$_1$ &  8.713 & 8.710  &  8.522  &  0.021  &  0.706 \\
 Ta$_2$ &  8.714 & 8.712  &  8.681  &--0.001  &--0.035 \\
 all Ta &        &        &  8.602  &  0.010  &  0.670 \\
 \hline
 O$_1$  &--6.450 &--6.472 &--6.315  &--0.051  &  1.293 \\
 O$_2$  &--6.529 &--6.504 &--6.413  &--0.034  &  0.867 \\
all O$_\parallel$ &  &    &--6.364  &--0.042  &  2.160 \\
 O$_3$  &--1.703 &--1.700 &--1.708  & -0.051  &  0.346 \\
 O$_4$  &--1.700 &--1.700 &--1.687  & -0.034  &  0.231 \\
 O$_5$  &--1.687 &--1.685 &--1.673  & -0.046  &  0.306 \\
 O$_6$  &--1.687 &--1.685 &--1.662  & -0.096  &  0.637 \\
all O$_\perp$  & &        &--1.683  & -0.057  &  1.521 \\
 all O  &        &        &         &         &  3.681 \\
 \hline
Total   &\par    & \par   &         &         &  5.357 \\
\multicolumn{2}{c}{Exact value}   & \par   &         &         &
5.153
\\
\end{tabular}\label{Berry}
\end{table}

The polarization induced by Li off-centering was studied by Berry
phase analyses, and dynamical charges $Z^*_{\alpha\alpha}$~ were
determined for all ions in KLT40$_{\rm Ref}$, fully relaxed KLT40
(Table \ref{DynCh}), and KLT40 with ions in ideal perovskite
positions (Table \ref{Berry}). Compare these charges with
$Z_{zz}$~ in KT with the same lattice parameter: K: 1.151; Ta:
8.680; O$x$: --1.690; O$y$: --1.690; O$z$: --6.452.

The dynamical charge is a symmetric function of displacements
($u$) from ideal perovskite positions, so to a first
approximation, it is quadratic in $u$:

\begin{equation}
Z^*_{zz}(u)=Z^*_{zz}(0)-\zeta u^2
\end{equation}

The average dynamical charge, which is required for calculating the
polarization, is

\begin{eqnarray}
\left<Z^*_{zz}\right>=\frac{1}{u_1-u_0}\int\limits_{u_0}^{u_1}Z^*_{zz}(u)du
= \nonumber\\ =
Z^*_{zz}(0)+\frac{1}{3}\zeta\left(u_1^2+u_0^2+u_1u_0 \right)
\end{eqnarray}

In particular, if $u_0=0$~ then

\begin{equation}
\left<Z^*_{zz}\right>=\frac{1}{3}\left(Z^*_{zz}(u_1)+2Z^*_{zz}(0)\right)
\end{equation}

Average values of the dynamical charges, and individual ion contributions
to the total KLT40 supercell dipole moment are listed in
Table~\ref{Berry}.  For comparison, the explicit value of the total dipole
moment, obtained by Berry phase analysis, is also shown.

The largest contribution to the total dipole moment is from O ions. The
Li-displacement contribution is roughly equal to that of Ta.
The total dipole moment is $\sim$5 times larger than the dipole
moment due to the Li ion itself, which indicates that the Li-dipole
moment is strongly \emph{enhanced} by structural relaxation.

A similar enhancement was also obtained in a shell model calculation,
\cite{Stachiotti} but in that study the main cause of enhancement was a
large Ta-displacement. Here, enhancement of the Li-dipole moment is
primarily caused by coupling between Li- and O-displacements.

\begin{table}[!htbp]
 \caption{The KLT80b supercell dipole moment (in e\AA) for different Li-Li
 nn-pair configurations}
\begin{tabular}{c|c|cccccc}
 notation & direct. & K & Li & Ta & O$_\perp$ & O$_\parallel$ & Total
 \\ \hline
 $z-z$            & z &  0.138 &  2.688 &  4.636 &  2.215 &  3.850 & 13.527 \\
 $x-\overline{z}$ & z &--0.170 &--1.168 &--3.702 &--1.432 &--2.524 &--8.996 \\
                  & x &  0.335 &  1.138 &  4.915 &  1.085 &  1.888 &  9.361 \\
 $x-x$            & x &  0.393 &  2.184 &  5.681 &  2.079 &  3.417 & 13.754 \\
 $x-z$            & z &  0.152 &  1.408 &  3.830 &  1.187 &  2.202 &  8.779 \\
                  & x &  0.512 &  1.193 &  6.315 &  0.580 &  1.008 &  9.608 \\
 $y-x$            & y &  0.393 &  1.147 &  5.451 &  0.867 &  1.480 &  9.338 \\
                  & x &  0.393 &  1.147 &  5.451 &  0.867 &  1.480 &  9.338 \\

 \end{tabular} \label{Li2dipmom} \end{table}

Total dipole moments for a KLT80$b$~ supercell with different
Li-Li nn-pair configurations are listed in Table~\ref{Li2dipmom}.
They were calculated with the average dynamical charges listed in
Table \ref{Berry} and ionic displacements relative to ideal
perovskite. Individual ionic contributions are also shown. Oxygen
ion contributions are separated into O$_\parallel$~ and O$_\perp$,
with the polarization direction parallel and perpendicular to the
Ta-O-Ta bond, respectively. These results indicate that the total
dipole moment induced by Li-Li pairs is larger than that from two
isolated Li-ions.

\subsection{Coupling Between Strain and Polarization}\label{stress}

The influence of applied stress on dipole moment and energy was
studied by applying various tetragonal distortions to the KLT40
supercell with $a=3.95594$~\AA, and $0.98a <c <1.02a$.  Total
energy calculations show that, for $c > a$, configurations with
$[0,0,1]_{c}$-displaced Li-ions are lower in energy than those
with $[1,0,0]_{c}$-displaced Li. The difference in energies is

\begin{equation}
-\delta E_{zz} [meV] = 1.649 e_3 + 105.7 e_3^2 + 3430. e_3^3 \label{Es}
\end{equation}

\noindent where $e_3 = (c/a-1)$. By using this energy difference,
the occupation probability for Li-displacements in the
$c$-direction can be estimated as

\begin{eqnarray}
w_{-z-z}=w_{zz} = \frac{ e^{-\delta E_{zz} / k_B T}}{4+2
e^{-\delta E_{zz} / k_B T}}\nonumber\\ \approx 1/6-\delta
E_{zz}/9k_B T
\end{eqnarray}
Hence, tetragonal strain increases the population of
Li-displacements parallel to the $c$-axis. The energy decrease
(per primitive cell) connected with redirected Li-ions is,

\begin{eqnarray}
<\delta E_{Lizzzz}>=x\delta E_{zz} (w_{-z-z} +
w_{zz}-1/3)\nonumber
\\ \approx -2 x \delta E_{zz}^2 /9k_B T
\end{eqnarray}
which at small $e_3$, is proportional to $e_3^2$. Thus, the energy
decreases by local Li-redirections along $c$~ and this decreases
the corresponding elastic constant proportionally to $x$.

Li-redirection also increases the square of supercell dipole
moment $d_z$~ along the $c$-axis

\begin{eqnarray}
\delta q_{zzzz} = d_{z}d_{z}(w_{zz}+w_{-z-z}-1/3)\nonumber
\\ \approx 2 d_{z}^2 \delta E_{zz}/9 k_B T.
\end{eqnarray}
By comparison with (\ref{Es}), $\delta q_{zzzz}$~ is proportional
to strain for small $e_3$. Because Li-redirection is coupled to
strain, one expects a relaxation contribution to the
electrostriction constants and acoustic response, as observed by
Pattnaik and Toulouse [\cite{Toulouse,Pattnaik}].

In addition to promoting Li-redirection, tetragonal distortions
change the TO soft mode frequency. Using the average dynamical
charges and computed ionic displacements, one can calculate
supercell dipole moments, $d_z$~ and $d_x$~ in $e$\AA, which
correspond to Li-ions displaced in the $[0,0,1]_{c}$~ and
$[1,0,0]_{c}$~ directions respectively, for $a<c<1.2a$:

\begin{eqnarray}
d_z = 5.15 + 167. e_3 + 2987. e_3^2\nonumber \\ d_x = 5.15 - 40.
e_3 + 585. e_3^2
\end{eqnarray}

These results show that the supercell dipole moment is greatly
enhanced by an increase in $c/a$. Partial contributions to
the dipole moment that are associated with Ta- and O-displacements
are most sensitive to changes in $c$; which is consistent with an
electrostriction induced reduction in the soft-mode frequency.

\subsection{Toward an Effective Hamiltonian}\label{Hamiltonian}

The picture of KLT in which Li-dipoles, a soft-mode, and PNC all
interact, lends itself to a first-principles based effective
Hamiltonian, $H_{eff}$, treatment.  Full formulation of $H_{eff}$~ is
beyond the scope of this work, but some essential intermediate
results have been derived. For simplicity, consider KLT40, with
the same ionic displacements in each unit cell.

To derive $H_{eff}$~ for a FE, one selects a local basis for the
instabilities that cause the FE transition, and includes interactions between
sites, between local distortion and strain, etc.\cite{Zho95b}.
In KLT, it is clear that instabilities centered on Li are responsible for
the phase transition, and that Li-centered distortion variables must
be included in $H_{eff}$. However, the coupling of Li-dipole moments
to the KT soft-mode is sufficiently important that one must also include
a local variable for the KT soft-mode.  By analogy with $H_{eff}$~ for
KNbO$_3$\cite{Wag98}, this local variable is centered on the Ta sites.

Several eigenvectors are relevant:

\begin{itemize}
\item{$v_{1,\alpha}$~: the dynamical matrix
eigenvector for the KT soft-mode, polarized along $\hat{\alpha}$.}

\item{$v_{2,\alpha}$~: the dynamical matrix eigenvector for the
Li-instability of KLT40, polarized along $\hat{\alpha}$.
(Dynamical matrix eigenvectors are normalized for 40-atom cells).}

\item{$v_{Ta,\alpha} = v_{1,\alpha}$ and
$v_{Li,\alpha} = v_{2,\alpha}$, orthogonalized to $v_{1,\alpha}$,
and renormalized.}

\item{$d_{Li,\alpha}$ and $d_{Ta,\alpha}$,
the dimensionless displacement eigenvectors obtained
by dividing the elements of $v_{Ta,\alpha}$ and $v_{Li,\alpha}$, respectively,
by $\sqrt{(m/m_0)}$, where (arbitrarily) mass $m_0$ = 1 amu.}

\end{itemize}

Consider displacement patterns in KLT40 in which displacement
amplitudes corresponding to $v_{Ta,\alpha}$~ and $v_{Li,\alpha}$~
are $\tau_{\alpha}$~ and $\lambda_{\alpha}$, respectively.  The
energy is minimized with respect to all other modes. The
strain tensor is $\{e_i\},~i=1,6$, in Voigt notation.
Calculated results closely fit the expansion:

\begin{eqnarray}
U  & = &  U _o \nonumber \\ & - &  0.055566 |\lambda|^2 + 0.004160
|\lambda|^4\nonumber\\ & + &  0.016303 (\lambda_x^2 \lambda_y^2 +
\lambda_x^2 \lambda_z^2 + \lambda_y^2 \lambda_z^2) \nonumber\\ & -
&  0.000077 |\lambda|^6  + 0.000001 |\lambda|^8 \nonumber\\ & + &
0.024760 |\tau|^2 + 0.008677 |\tau|^4 \nonumber\\ & - &  0.011288
(\tau_x^2 \tau_y^2 + \tau_x^2 \tau_z^2 + \tau_y^2 \tau_z^2)
     + 0.000684 |\tau|^6  \nonumber \\
& - &  0.025236 \vec{\lambda} \cdot \vec{\tau}
     + 0.001210 (\lambda_x^3 \tau_x + \lambda_y^3 \tau_y + \lambda_z^3 \tau_z) \nonumber \\
& - &  0.000370 (\lambda_x \tau_x (\lambda_y^2 + \lambda_z^2) +
            {\rm cyclic~perm.})  \nonumber \\
& + & 817.342 (e_1^2 + e_2^2+ e_3^2) + \nonumber \\&+&210.580 (e_1
e_2 + e_1 e_3 + e_2 e_3) \nonumber \\ & + & 732.803 (e_4^2 + e_5^2
+ e_6^2) \nonumber \\ & - & 0.262724 (e_1 \lambda_x^2 + e_2
\lambda_y^2 + e_3 \lambda_z^2) \nonumber \\ & - & 0.209816 (e_1
(\lambda_y^2 + \lambda_z^2) + e_2 (\lambda_x^2 + \lambda_z^2)+
\nonumber \\ &+&
              e_3 (\lambda_x^2 + \lambda_y^2))                \nonumber\\
& - & 5.30307 (e_4 \lambda_y \lambda_z + e_5 \lambda_x \lambda_z +
e_6  \lambda_x \lambda_y) \nonumber\\ & - & 3.24098 (e_1 \tau_x^2
+ e_2 \tau_y^2 + e_3 \tau_z^2) \nonumber\\ & + & 0.715628 (e_1
(\tau_y^2 + \tau_z^2) + e_2 (\tau_x^2 + \tau_z^2) + \nonumber \\
             &+& e_3 (\tau_x^2 + \tau_y^2))                \nonumber\\
& + & 1.38297 (e_4 \tau_y \tau_z + e_5 \tau_x \tau_z + e_6  \tau_x \tau_y), \nonumber\\
\label{heff.eq}
\end{eqnarray}
where U is in eV and $\vec{\lambda}$~ and $\vec{\tau}$~ in~\AA.
The connection between TO polarization and $\tau$~ is,
$P_{TOz}=3.766820\tau_z/V$~ $e$\AA~ where $V$~ is the unit cell
volume. The connection between the Li-dipole moment and
$\vec{\lambda}$~ is, $\mu_z=0.876855 \lambda_z$~ $e$\AA.

The potential for $\vec{\lambda}$~ alone ($\sim$~ Li off-centering)
has 6 wells in $[0,0,1]_{c}$~ directions. The potential for
$\vec{\tau}$~ alone (KT soft-mode) has a small positive
harmonic coefficient and is therefore highly polarizable. Negative
bilinear coupling between $\vec{\lambda}$~ and $\vec{\tau}$~ indicates
that the soft-mode enhances the total polarization from Li
off-centering.  Positive coefficients of $|\tau|^4$~
and $|\tau|^6$ imply that the soft-mode frequency hardens as the
structure distorts.  Higher-order coupling between
$\vec{\lambda}$~ and $\vec{\tau}$~ is included because it significantly
improves the fit.

One can reduce the number of variables by excluding strain. This
results in a decrease of the nonlinearity constant which
multiplies the fourth power of polarization, and changes the sign
of the angle dependent term for the soft-mode.

Eq.~\ref{heff.eq} and the mode effective charges can be used to
calculate the soft-mode and Li-dipole contributions to the
susceptibility.  The soft-mode contribution is 110
for the reference state, much less than might be
expected considering that KT is an incipient ferroelectric.
The LDA overestimates the soft mode frequency of KT,
which leads to too large a value for the $|\tau|^2$~ coefficient of
Eq.~\ref{heff.eq} and too low a permittivity.
For fully relaxed KLT40, the model predicts that the
contribution of the soft-mode to the susceptibility along
the polar axis is about 55 due to mode hardening.
The Li-dipole contribution is insignificant (order 1).
This, however, is only the contribution to the
static susceptibility from Li vibrating within a single well.
As discussed in Section~\ref{Phenomenology}, Li-hopping
gives a much greater, frequency-dependent,
contribution to the susceptibility.

From Eq.~\ref{heff.eq} and the mode effective charges, the
coupling between TO phonons and Li-dipoles is $0.064 \mu_z
P_{TOz}/3V\varepsilon_0$~ where $\varepsilon_0$~ is the
permittivity of free space. This coupling enhances dielectric
permittivity and shifts the Curie temperature to higher values
(c.f. Sec \ref{Phenomenology}). This Lorentz factor, 0.064, is
smaller than that found \cite{Vug90} in a
point polarizable ion model, (0.1 - 0.2). The Lorenz factor
decreases quadratically with Li displacement and equals 0.037 at
the equilibrium Li position.

To complete $H_{eff}$~ requires:
(1) Formulating local representations of $\vec{\lambda}$~ and $\vec{\tau}$~
e.g. lattice Wannier functions\cite{Rab95}); (2) Quantifying
distance dependencies of the interactions\cite{Zho95b, Wag97}.

\subsection{Band Structure Calculations}

Strong n-type photocurrent is observed in KLT at $T \lesssim ~80
K$ \cite{Glinchuk}, and it has been attributed to shallow O$_62p$~
hole states at the bottom of the valence band which are caused by
large O$_6$-displacements (notation of Table \ref{POSCAR}). Holes
are trapped by these levels which prevent recombination with
electrons that were promoted to the conduction band by light
absorption.

To study shallow forbidden gap states in KLT, band-structure
calculations were performed for KLT40 and the corresponding ideal
KTaO$_3$~ supercell, in the $\Gamma - X$~ direction. Results for
the top of valence and bottom of conduction bands are shown in
Fig. 10. The top of the valence band, and the bottom of the
conduction band, are split and/or distorted in KLT40 relative to
KT.

VASP results confirm large O$_6$-displacements, \cite{Glinchuk}
and demonstrate some splitting at the top of the valence band, and
some distortion at the bottom of the conduction band. This occurs
because Li off-centering breaks Pm$\overline{3}$m symmetry
creating symmetrically different O-ions with different
self-consistent electrostatic potentials (cf. Ref.
\cite{domains}). The VASP results, however, do not
confirm the presence of O$_62p$~ states at the top of the
valence band.

\begin{table}[!htbp]
  \caption{Electrostatic potentials (in eV) on the
  ions of the KLT40 supercell in the equilibrium structure (Equil),
  reference structure (Refer), and in pure KT (KT).
  Only symmetrically distinct ions are listed (atomic positions
  are listed in Table \ref{POSCAR}. }
\begin{tabular}{c|ccc|c|ccc}

 Ion \par & Equil.&Refer.& KT&Ion \par & Equil.&Refer.& KT  \\
\hline
 K$_1$  &  10.15  &  10.18  &  10.33  &
 K$_2$  &  10.13  &  10.15  &  10.33  \\
 K$_3$  &  10.18  &  10.23  &  10.33  &
 K$_4$  &  10.15  &  10.15  &  10.33  \\
 K$_5$  &  10.14  &  10.23  &  10.33  &
 Li     &  43.90  &  42.68  &  \par    \\
 Ta$_1$ & --1.21  & --1.22  & --1.30  &
 Ta$_2$ & --1.29  & --1.22  & --1.30  \\
 O$_1$  &--56.27  &--56.26  &--56.07  &
 O$_2$  &--56.30  &--56.30  &--56.07  \\
 O$_3$  &--56.19  &--56.26  &--56.07  &
 O$_4$  &--56.46  &--56.26  &--56.07  \\
 O$_5$  &--56.28  &--56.31  &--56.07  &
 O$_6$  &--56.43  &--56.31  &--56.07  \\
\end{tabular}\label{Potentials}
\end{table}

Computing the electrostatic potentials on different ions in KLT40
(Table \ref{Potentials}) reveals that the O$_62p$~ states are
significantly deeper in fully relaxed KLT40 than in KT or
KLT40$_{\rm Ref}$. The same result was obtained from projections
of band states onto O$_62p$~ states:  O$_62p$-states in KLT40 lie
below the top of the valence band. Similar results were obtained
for Li-Li nn pairs in KLT80, where the tetragonal distortion from
Li also causes valence and conduction band splitting.

\section{Phenomenology}\label{Phenomenology}
\subsection{Li-pair relaxation time}

 Consider KLT as a distribution of PNC (containing
one or more Li-ions) in a KT lattice.  FP calculations
on isolated Li-ions and on Li-Li pairs have shown that each
PNC may have a set of states defined by the Li-displacement
directions. For a given PNC, group theory can be used to
enumerate distinct relaxation processes and identify which
of them couple to an electric field.  The kinetic equations for
state occupation probabilities can be solved to estimate
the relaxation times.  If PNC do not interact, then
there are distinct relaxation times for each cluster type,
and the relaxational contribution to the dielectric function
is the sum of contributions from each cluster.
Interactions between clusters will broaden an otherwise
sharp distribution of relaxation times.

For isolated Li-ions, there are six states, and
symmetry analysis of relaxation processes yields $A_{1g}$~ + $E_g$~ +
$F_{1u}$.  The $A_{1g}$~ ``process" is the trivial steady
state equilibrium solution.
Only the $F_{1u}$~ process couples with an electric field and
affects dielectric dispersion.  The $E_g$~ process couples with stress,
but not electric field.  Let $k$~ be the hopping rate between wells
separated by $90^{\circ}$~ and neglect hopping between wells separated
by $180^{\circ}$.  Then relaxation time $\tau^{-1}=4k$~ for the $F_{1u}$~
and $\tau^{-1}=6k$~ for the $E_g$~ process\cite{Borsa}.

For Li-Li nn-pairs, there are 36 states, many of which are
metastable (Table~\ref{Li2excitations}).  Symmetry analysis
identifies seven relaxation processes that contribute to the dielectric
function.  Six of these involve substantial probability
redistribution among metastable states whose occupancies are
negligible.  Thus, only one relaxation process contributes
significantly to the dielectric function; the one whose
net effect is $z-z\rightleftharpoons \overline{z}-\overline{z}$.

The full set of kinetic equations give a $36 \times 36$~ matrix.
For simplicity, consider only the states that are associated with
the minimum energy path in Fig. 7:

\begin{eqnarray}
 &&-\frac{dw_a}{dt}=k_{ab}w_a-k_{ba}w_b
 \nonumber \\
 &&-\frac{dw_b}{dt}=(k_{ba}+k_{bc})w_b-k_{ab}w_a-k_{cb}w_c
 \nonumber \\
 &&-\frac{dw_c}{dt}=2k_{cb}w_c-k_{bc}w_b-k_{bc}w_d
 \nonumber \\
 &&-\frac{dw_d}{dt}=(k_{bc}+k_{ba})w_d-k_{cb}w_c-k_{ab}w_e
 \nonumber \\
 &&-\frac{dw_e}{dt}=k_{ab}w_e-k_{ba}w_d
\end{eqnarray}
where $w_\alpha$~ is occupation probability for well $\alpha$. The
relaxation time of the dominant process is:

\begin{eqnarray}
\tau^{-1}=(k_{ab}+k_{ba}+k_{bc})/2 -
\sqrt{a^2-k_{ab}k_{ba}},
\nonumber \\
\end{eqnarray}

\noindent Here $a = (k_{ab}+k_{ba}+k_{bc})/2$.

In the realistic limit
$k_{ab} \sim k_{bc} \ll k_{cb} \sim k_{ba}$: $ \tau^{-1} \simeq
k_{bc}k_{ab}/k_{ba}$.  This rate
corresponds to overcoming the principle Li-Li
nn-pair barrier ($z-z \rightarrow x-x$) in the $z-z
\rightleftharpoons \overline{z}-\overline{z}$~ process.

\subsection{Dielectric-permittivity frequency-dependence}

If one excludes strains from a mean field Hamiltonian then there
are two variables: one for a soft-mode; the other for Li-dipoles.
Both contribute to the dielectric response, but the former is
essentially dispersionless far below the soft-mode frequency. The
Li-dipole term however has relaxational character. Neglecting the
damping of Li-oscillations in Li wells, but accounting for
jumps between nn Li-wells, the Li-dipole associated with
$z$-polarization of an isolated Li-ion in a cubic environment is:

\begin{equation}
D_z=\frac{x \mu_z}{V} \left( w_z-w_{-z} \right)
\end{equation}

\noindent where $V$~ is the primitive unit cell volume. The
$F_{1u}$ relaxation process contributes to polarization
fluctuations, and the polarization takes the form \cite{FTT}

\begin{equation}
D_z= \varepsilon_0 F(T) \tilde{E}_z
\end{equation}

\noindent where

\begin{equation}
F(T)=\frac{x \mu_z^2}{V\varepsilon_0}\frac{1}{6k_B T} \frac{1}{4k
+ i \omega} \label{F(T)}
\end{equation}

\noindent Here: $\tilde{E}_z = E_z + \lambda P_{TOz}$~is the local
field, $\lambda=\lambda_0 - 3\lambda_1 \mu^2$~ a coupling
constant, $\lambda_0$~ and $\lambda_1$~ are the effective
Hamiltonian coefficients of $-\mu_z P_{TOz}$~ and $\mu_z^3
P_{TOz}$~ respectively; $E_z$~ is an external field; $P_{TOz}$~ is
the polarization connected with the soft-mode; and $\tau =4k$~ is
the relaxation time for the $F_{1u}$ relaxation process. The
dielectric susceptibility connected with Li jumps can be now
written as \cite{Vug90,Su,FTT}

\begin{eqnarray}
 \chi_{Li}=
 \frac{F(T)}
 {1-\lambda^2F(T)\varepsilon_0/ A(T) }
\end{eqnarray}
where

\begin{equation}
A(T)=\alpha +3 \beta P_{TOz}^2 + 5 \gamma P_{TOz}^4
\end{equation}

\noindent Here the nonlinearity coefficient $\beta$~ is
renormalized after the elimination of strains; $\alpha$~ is
renormalized due to fields produced by Li.  The positive
correlation between $\alpha$~ and increased $x$~ implies hardening
of the soft mode with increasing Li concentration. Dielectric
susceptibility is enhanced at low frequencies by coupling between
Li-dipoles and the TO soft mode; $T_{c}$~ increases and frequency
dispersion is enhanced. These results were obtained by considering
all hopping paths between the six $[0,0,d]_{c}$-type wells, not by
invoking an Ising model.

The dominant relaxation process for Li-Li nn pairs leads to an
additional relaxational contribution to $F(T)$, proportional to
Li-Li nn-pair concentration, and with $\tau^{-1} =
k_{bc}k_{ab}/k_{ba}$ (cf. results in Refs. \cite{Trepakov,JETP}).
The concentration of Li-Li nn pairs is low, but their dipole
moments are large (Table \ref{Li2dipmom}), and $\chi_{Li}$~ is
proportional to the dipole moment squared (\ref{F(T)}).

The temperature dependence of the kinetic coefficient $k$~ is
common, and it is mainly controlled by the barrier height $U$. The
temperature dependence of the relaxation time is

\begin{eqnarray}
\label{eqtau} \tau &&= A\chi(T)\int\limits_{ - \infty }^\infty
{e^{\left( {U + ay^2} \right) / k_B T}dy} = \nonumber \\
&&=A\chi(T)\sqrt {\frac{\pi k_B T}{a}} e^{U\left( T \right) / k_B
T} = \nonumber \\ &&= \tau _0 \left( T \right)e^{U\left( T \right)
/ k_B T}, \label{saddle}
\end{eqnarray}
where $\chi$~ is the dielectric permittivity of the media in which
the PNC are embedded and $a$~ is a constant determined by the
force constants at the saddle point.

\section{Conclusions}

Calculations on a 40-atom KLT40 supercell that contains one Li-ion
indicate that there is only one instability connected with
Li off-centering. This instability is associated with large
Li-displacement in the $[0,0,1]_{c}$ direction.
Li potential wells are separated by potential
barriers with $[1,1,0]_{c}$ saddles. Calculated $[1,1,0]_{c}$~ potential
barriers and $[0,0,d]_{c}$~ Li-displacements, in fully relaxed KLT, are
in good agreement with values obtained by fitting a mean field
expansion to experimental data \cite{Toulouse,Trepakov}.

All ions in the KLT40 supercell are displaced from ideal
perovskite positions, but the O-ions closest to Li exhibit the
largest sympathetic displacements, towards Li.  Calculated
vibrational frequencies for fully relaxed KLT40 indicate that
$A_1$- and $E$-symmetry modes emanate from an $F_{1u}$~ TO mode
similar to the soft mode of pure KT.  The average frequency of
these modes is higher than that of the TO soft mode, which
reflects hardening of the soft-mode induced by substituting Li for
K. The intrawell Li vibrational frequency is about 300 cm$^{-1}$.
The soft-mode and Li-polarization are coupled.

Berry phase analyses yielded total dipole moments associated with
Li off-centering that are $\sim$5 times larger than the Li-dipole
moment itself. This enhancement is primarily caused by
O-relaxation around the off-centered Li-ion.

Fully relaxed KLT40 is tetragonal with $c/a > 1$, and
polarization along the c-axis.  Polarization increases
with increasing $c/a$; uniaxial strain.

First-principle calculations of Li-Li nn-pair excitations in a
KLT80 supercell strongly support the conclusion of Dousenau et al.
\cite{Dousenau} that the ``$180^\circ$" relaxation process is caused
by $z-z \rightleftharpoons \overline{z}-\overline{z}$~ Li-Li
nn-pair flipping.

\section*{Acknowledgment}

 We thank G. Kresse for providing an LDA pseudopotential for Ta.
 S.A.P. appreciates support from
RFBR grant \# 01-02-16029 and thanks NIST for hospitality.

\section*{ Captions}

Fig. 1. KLT configurations compared on a common
320-atom supercell. The grid connects perovskite A-sites.
Empty sites represent K; black circles represent Li.
When there is one Li per primitive cell, it is arbitrarily displaced
in the $\hat{z}$~ direction; when there are two Li per
primitive cell, one is displaced in the $\hat{z}$~
direction and the other in the $\hat{x}$ direction.

Fig. 2. Structural distortions around Li in KLT40.

Fig. 3. Energy as a function of Li-displacement from an ideal
A-site position in KLT40: with all ions' coordinates frozen in the
centrosymmetric positions except for Li (a); with full relaxation
of all ions' coordinates except for Li (b).

Fig. 4. Comparison of the reflectivity spectra computed for KT and
KLT40 (the component along the polar axis ($R_z$), perpendicular
to it ($R_x$), and averaged spectrum ($R_{av}$)) at identical
lattice parameters $a_{0}$ = 3.956~\AA. The electronic permittivity
$\varepsilon_{\infty}$ was set to 5.15 in each case, the
experimental value for KT.\cite{Klink} The damping constant was
set to 20 cm$^{-1}$~ for all modes in KT and 40 cm$^{-1}$~ for
KLT40.

Fig. 5. Li potential barriers in KLT40 in the direction between
nearest neighbor Li-wells computed with all ions except Li fully relaxed
(relaxed), and with all ions except Li frozen at their coordinates
corresponding to Li at its equilibrium position (frozen).

Fig. 6. A-site configurations in KLT80$b$. Undecorated sites are
occupied by K.

Fig. 7. The energetics of Li-Li nearest neighbor pair
reorientation, $z-z \rightarrow \overline{z}-\overline{z}$,
including the intermediate states depicted in Fig. 6.

Fig. 8. Densities per A-site of isolated Li-ions and Li-Li
nearest-neighbor pairs in K$_{1-x}$Li$_x$TaO$_3$, as functions of
$x$.

Fig. 9. Fourth nearest neighbor Li-Li configurations: a)
$(z-z_4)$; b) $(z-\overline{z}_4)$, and c) $ (z-x_4)$.

Fig. 10. Comparison of the electronic band structures in KT40 (a),
KLT40 (b), and in KLT80 with a $z - z$~ Li nn-pair (c). The
valence states are below zero energy.

\begin{widetext}

\begin{figure}
\resizebox{0.5\textwidth}{!}{\includegraphics{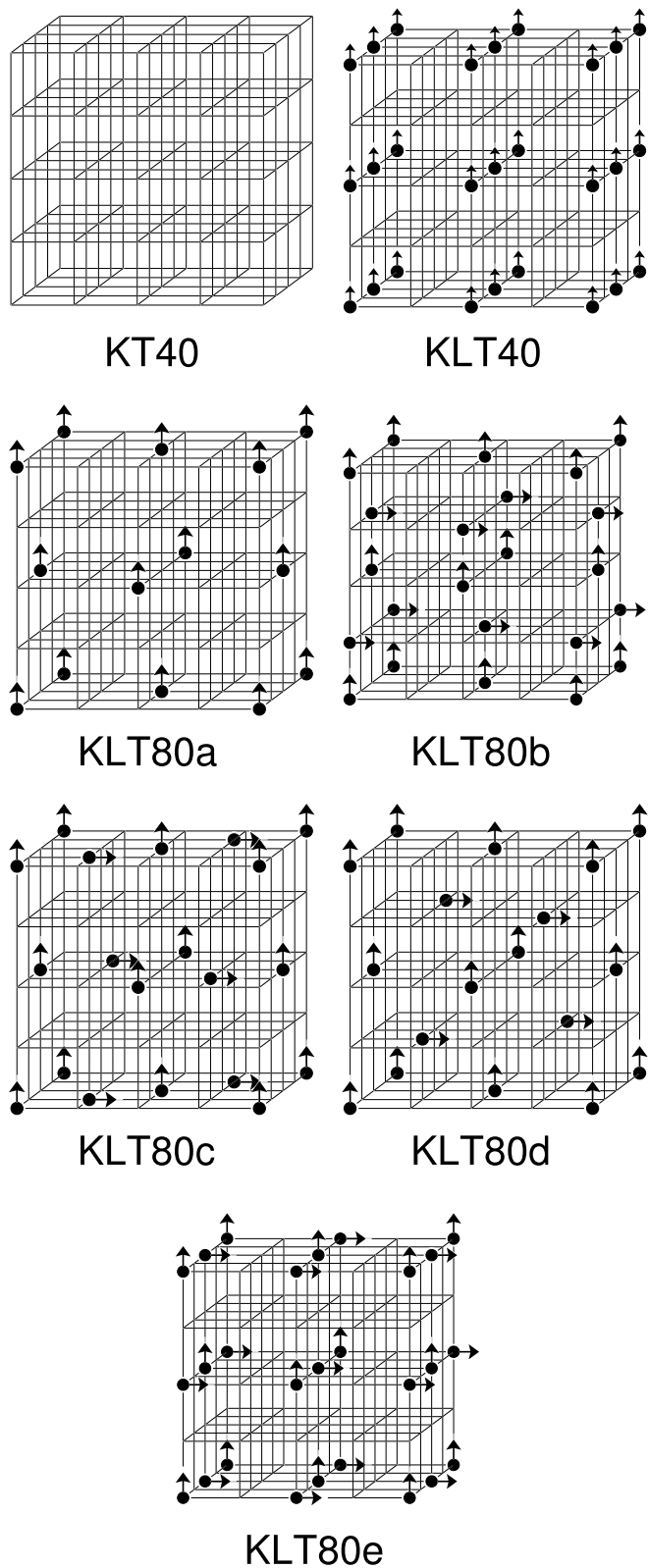}}
\caption{Prosandeev, Cockayne, Burton, PRB}
\end{figure}

\newpage

\begin{figure}
\resizebox{0.5\textwidth}{!}{\includegraphics{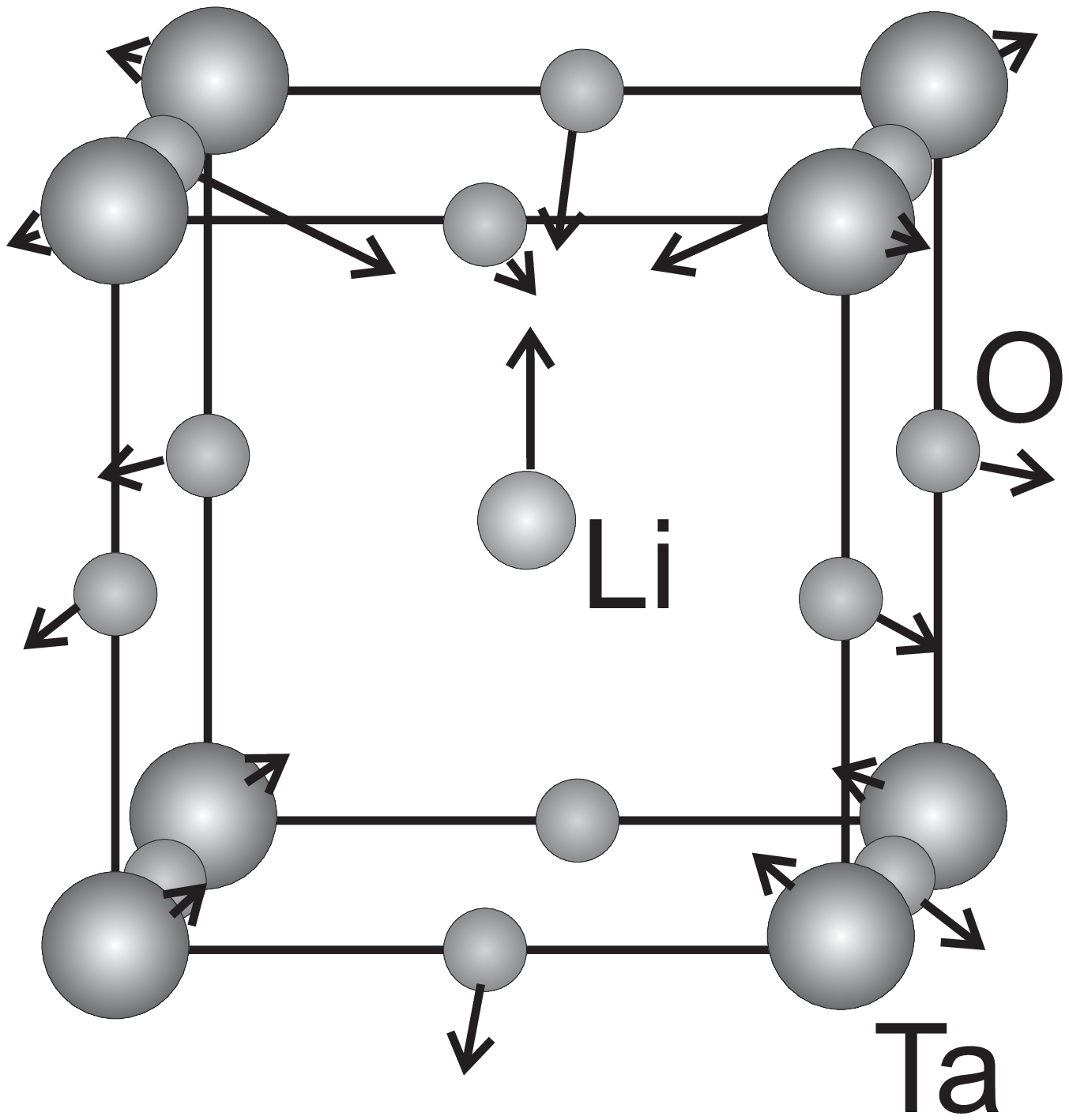}}
\caption{Prosandeev, Cockayne, Burton, PRB}
\end{figure}

\newpage

\begin{figure}
\resizebox{1.0\textwidth}{!}{\includegraphics{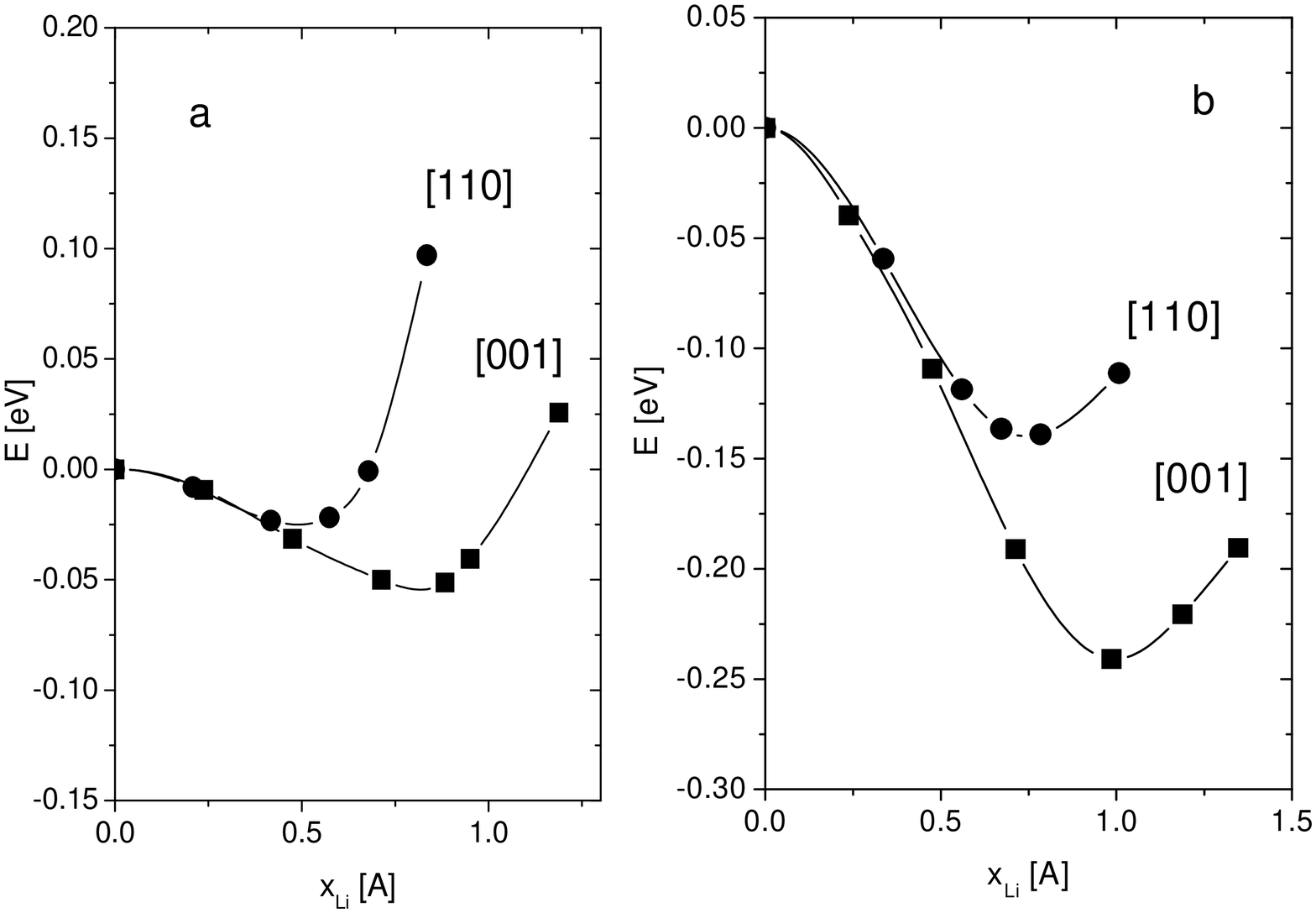}}
\caption{Prosandeev, Cockayne, Burton, PRB}
\end{figure}

\newpage

\begin{figure}
\resizebox{0.9\textwidth}{!}{\includegraphics{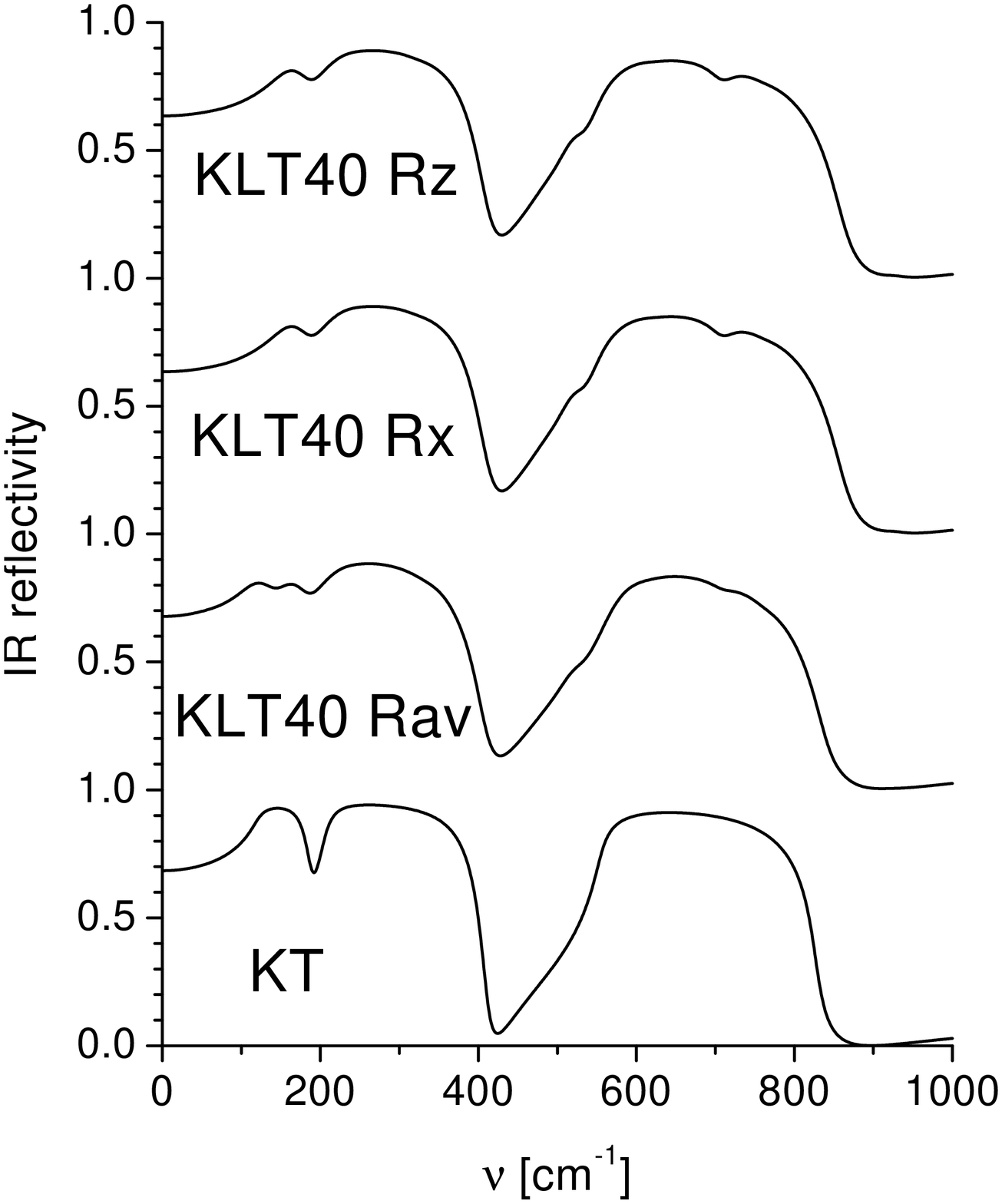}}
\caption{Prosandeev, Cockayne, Burton, PRB}
\end{figure}

\newpage

\begin{figure}
\resizebox{0.85\textwidth}{!}{\includegraphics{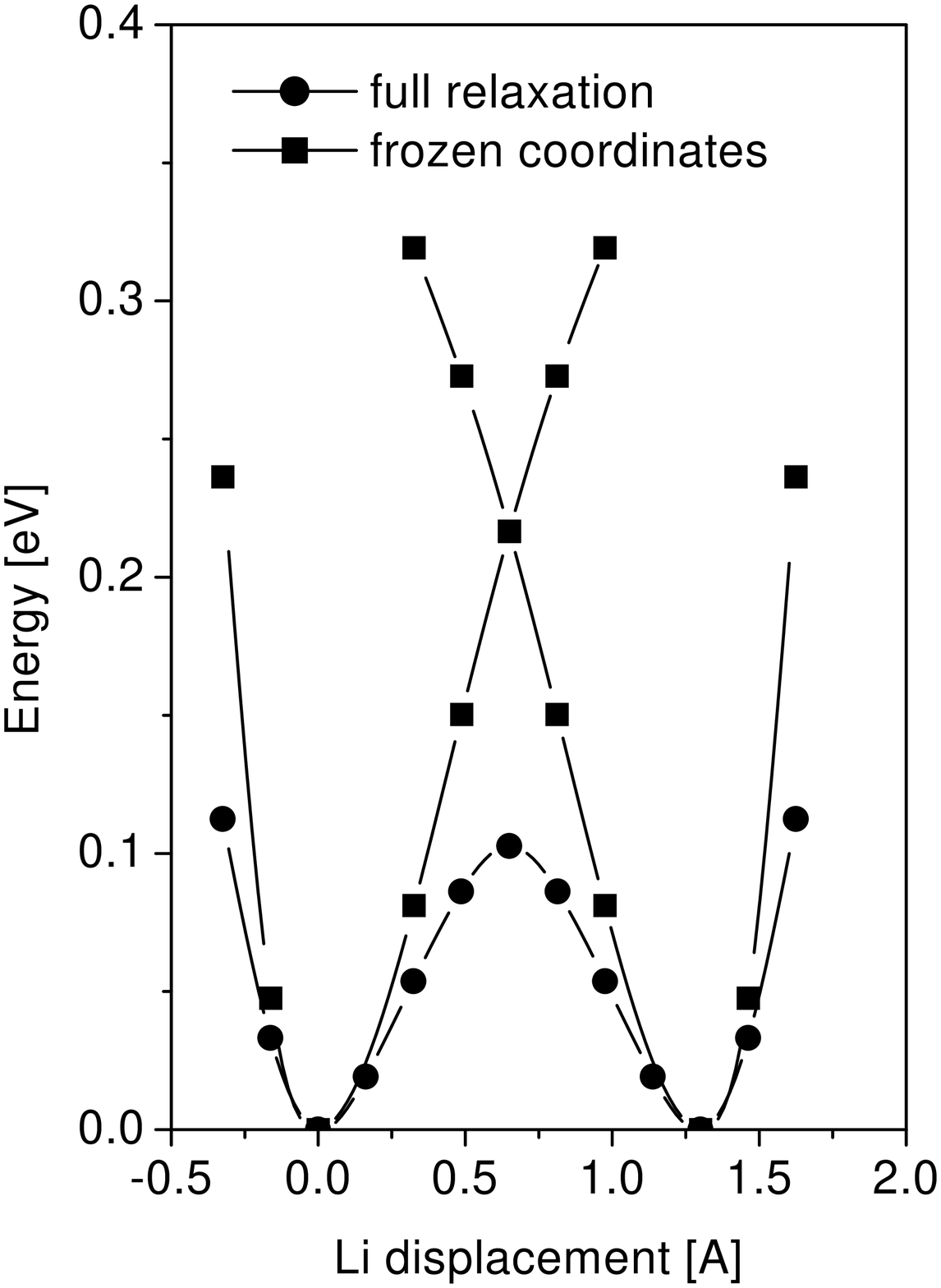}}
\caption{Prosandeev, Cockayne, Burton, PRB}
\end{figure}

\newpage

\begin{figure}
\resizebox{0.9\textwidth}{!}{\includegraphics{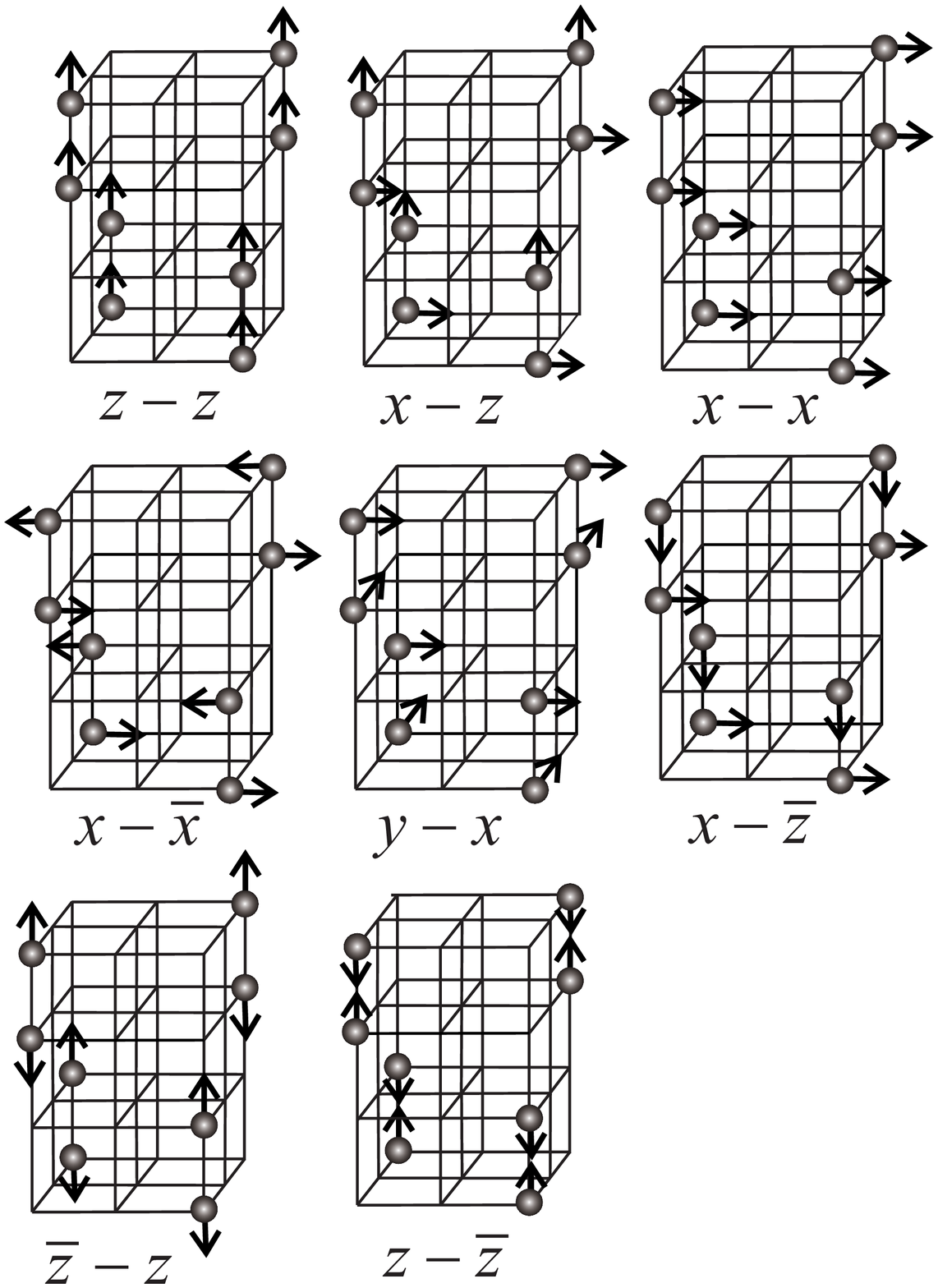}}
\caption{Prosandeev, Cockayne, Burton, PRB}
\end{figure}

\newpage

\begin{figure}
\resizebox{1.0\textwidth}{!}{\includegraphics{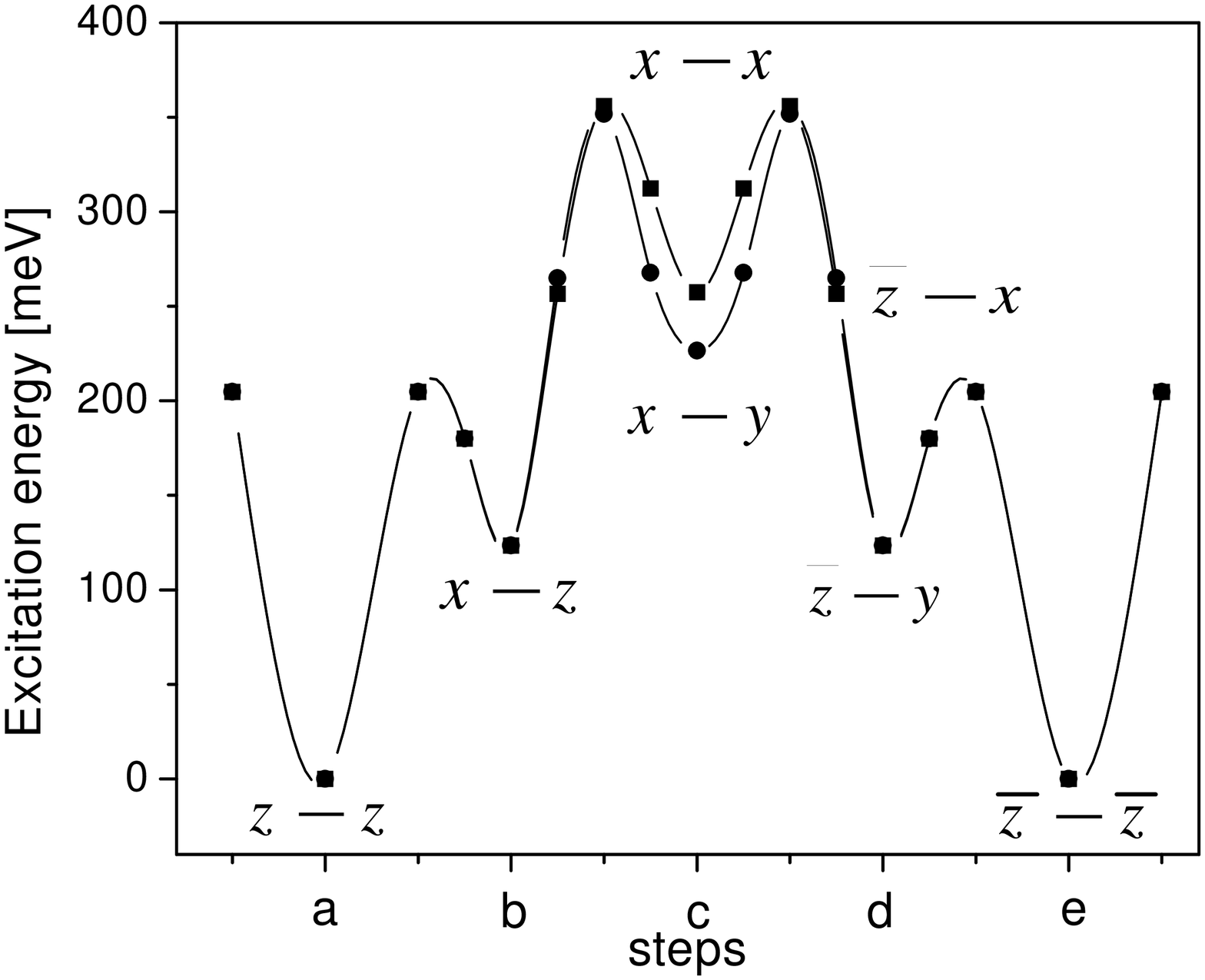}}
\caption{Prosandeev, Cockayne, Burton, PRB}
\end{figure}

\newpage

\begin{figure}
\resizebox{1.0\textwidth}{!}{\includegraphics{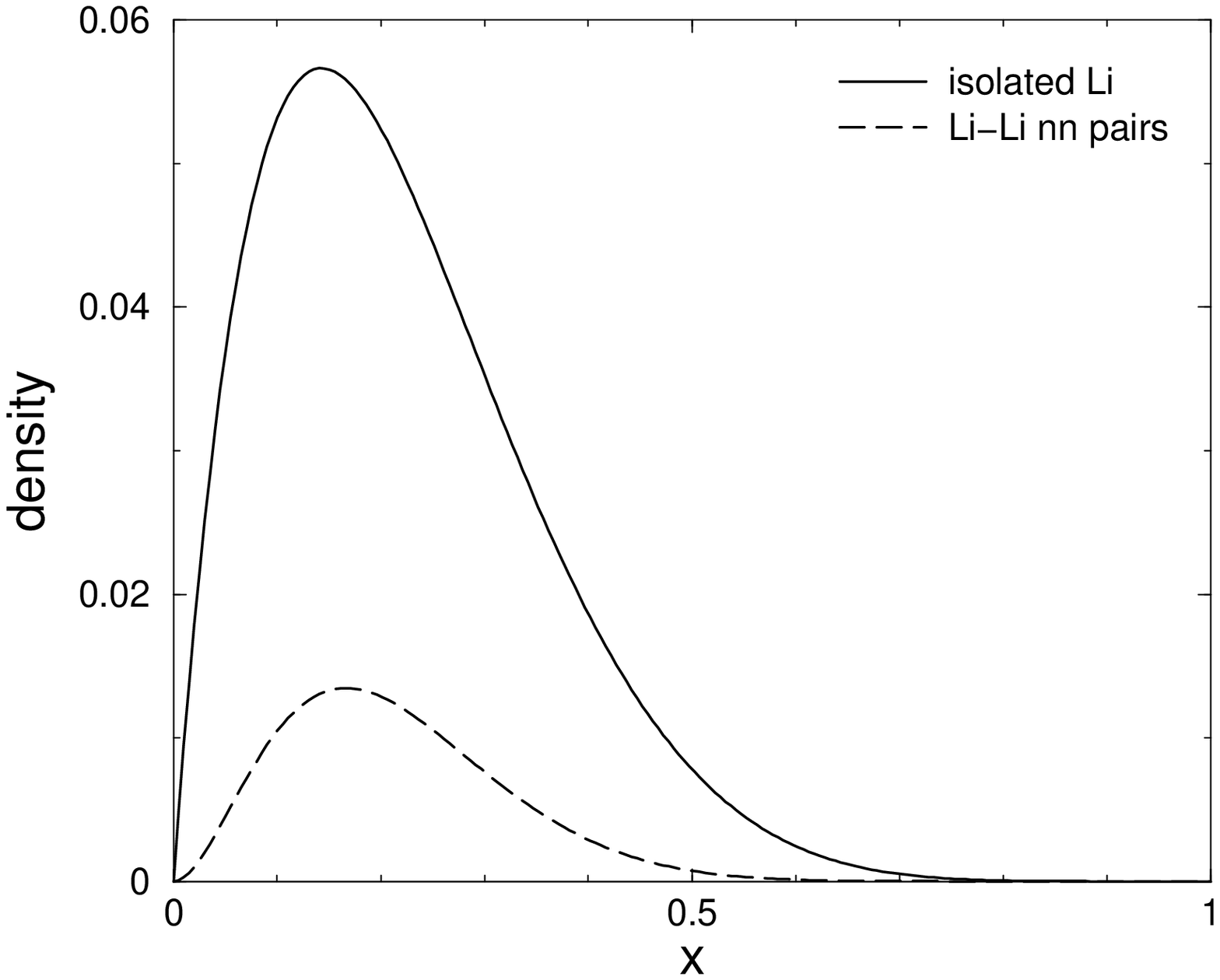}}
\caption{Prosandeev, Cockayne, Burton, PRB}
\end{figure}

\newpage

\begin{figure}
\resizebox{1.0\textwidth}{!}{\includegraphics{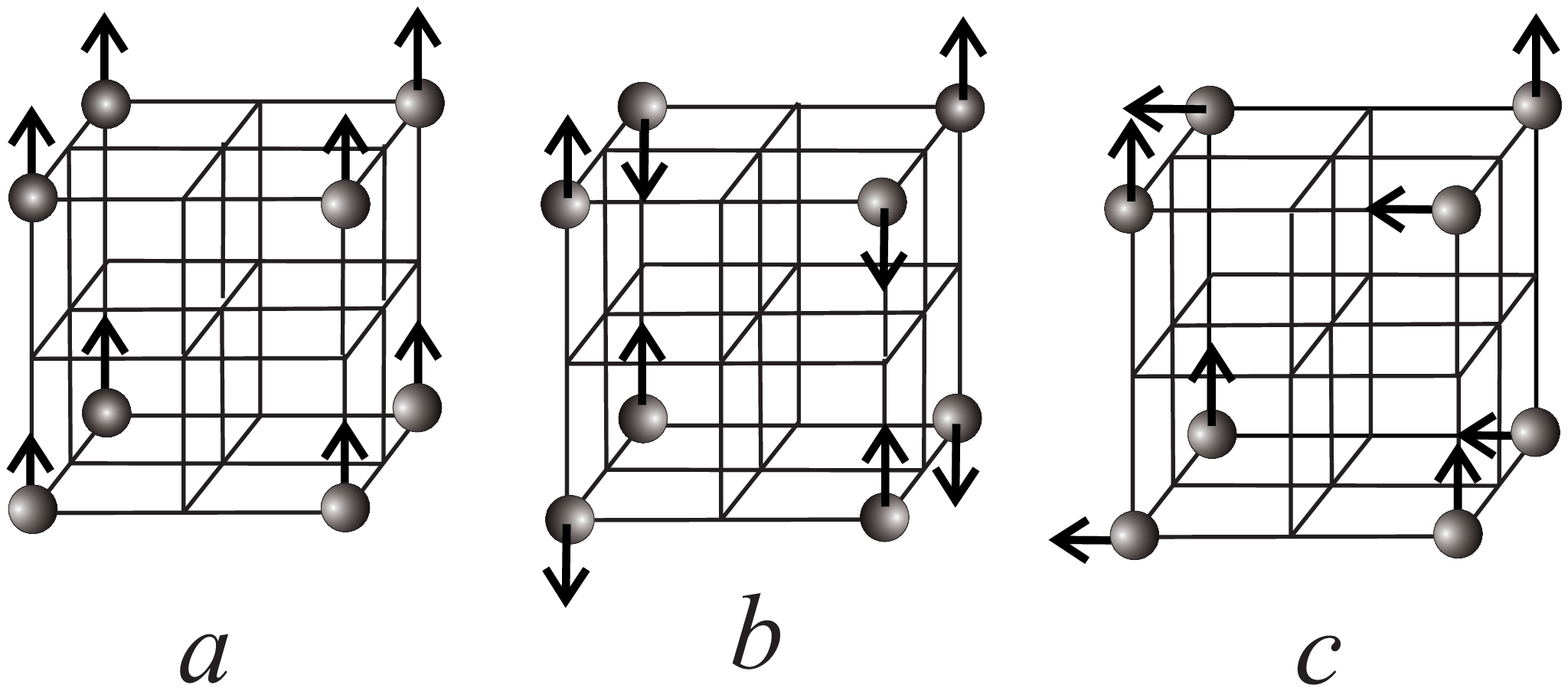}}
\caption{Prosandeev, Cockayne, Burton, PRB}
\end{figure}

\newpage

\begin{figure}
\resizebox{1.0\textwidth}{!}{\includegraphics{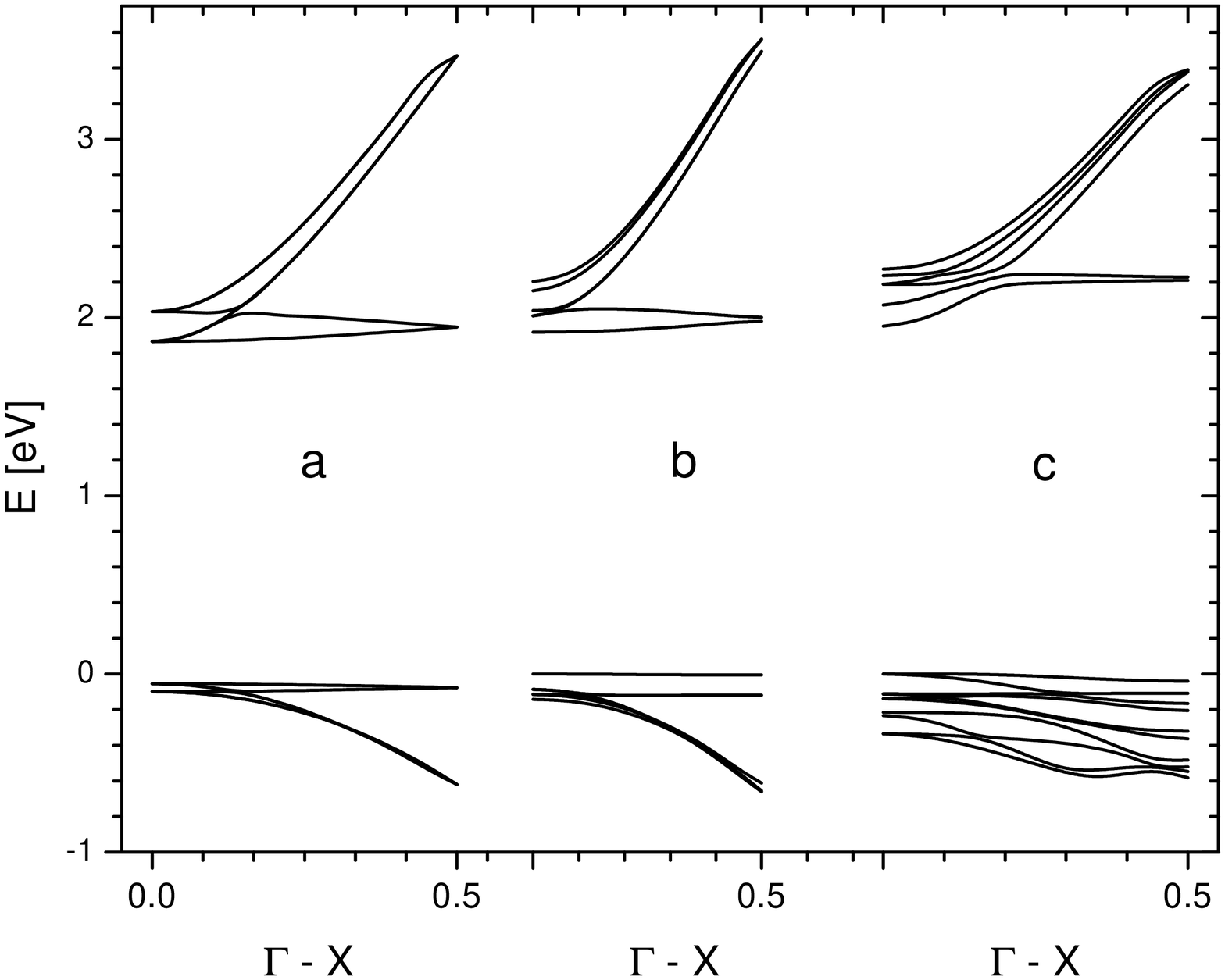}}
\caption{Prosandeev, Cockayne, Burton, PRB}
\end{figure}

\end{widetext}


\begin{thebibliography}{999}
\bibitem{foot1} Because of weak coupling with another mode
with similar frequency, the 205 cm$^{-1}$ mode of pure KT
actually is split into 200 cm$^{-1}$ and 205 cm$^{-1}$ modes of
KLT40.
\bibitem{foot2} All results in this paper are calculated
using periodic boundary conditions.  The results for
Li-displacements, etc., include interactions with a periodic
set(s) of image Li-ions and do not correspond to the limit
of a single Li-ion (single Li-Li pair) in an infinite
KTaO$_3$ lattice.
\end{thebibliography}
\end{document}